\documentclass[preprint]{aastex}

\newcommand{\ls}
 {\mathrel{\hbox{\rlap{\hbox{\lower4pt\hbox{$\sim$}}}\hbox{$<$}}}}
\newcommand{\gs}
 {\mathrel{\hbox{\rlap{\hbox{\lower4pt\hbox{$\sim$}}}\hbox{$>$}}}}

\def\la{\mathrel{\hbox{\rlap{\hbox{\lower4pt\hbox{$\sim$}}}{\raise2pt\hbox{$<$}}
}}}
\def\ga{\mathrel{\hbox{\rlap{\hbox{\lower4pt\hbox{$\sim$}}}{\raise2pt\hbox{$>$}}
}}}

\usepackage{amsmath}

\begin{document}

\title{%
 Investigating the Potential Dilution of the Metal Content of Hot Gas in Early-Type Galaxies by Accreted Cold Gas 
}%

\author{Yuanyuan Su\altaffilmark{1}
and
Jimmy A. Irwin\altaffilmark{1}
}

\altaffiltext{1}{Department of Physics and Astronomy, University of 
Alabama, Box 870324, Tuscaloosa, AL 35487, USA}

\email{ysu@crimson.ua.edu}

\begin{abstract}
The measured emission-weighted metal abundance of the hot gas in early-type
galaxies has been known to be lower than theoretical
expectations for 20 years. 
In addition, both X-ray luminosity and metal abundance vary significantly among galaxies of similar optical luminosities. This suggests some missing factors in the galaxy evolution process, especially the metal enrichment process.
With {\it Chandra} and {\it XMM-Newton}, we studied 32
early-type galaxies (kT $\lesssim$ 1 keV) covering a span of two orders
of $L_{X,\rm gas}/L_{K}$ to investigate these missing factors.  
Contrary to previous studies that X-ray faint galaxies show extremely low Fe abundance ($\sim 0.1$ Z${_\odot}$), nearly all galaxies in our sample show an Fe abundance at least 0.3 Z${_\odot}$, although the measured Fe abundance difference between X-ray faint and X-ray bright galaxies remains remarkable.  We investigated whether this dichotomy of hot gas Fe abundances can be related to the dilution of hot gas by mixing with cold gas.
With a subset of 24 galaxies in this sample, we find that there is virtually no correlation between hot gas Fe abundances and their atomic gas content, which disproves the scenario that the low metal abundance of X-ray faint galaxies might be a result of  the dilution of the remaining hot gas by pristine atomic gas. In contrast,  we demonstrate a negative correlation between the measured hot gas Fe abundance and the ratio of molecular gas mass to hot gas mass, although it is unclear what is responsible for this apparent anti-correlation. We discuss several possibilities including that externally originated molecular gas might be able to dilute the hot gas metal content.   
Alternatively, the measured hot gas Fe abundance may be underestimated due to more complex temperature and abundance structures and even a two-temperature model might be insufficient to reflect the true value of the emission weighted mean Fe abundance. 
\end{abstract}

\keywords{
galaxies: luminosities and abundance --
galaxies: elliptical and lenticular --
galaxies: ISM --
X-rays: galaxies
}

\section{\bf Introduction}
Most early-type galaxies (elliptical and lenticular galaxies) are composed of old, low-mass stars with little on going star formation activity.
Over several gigayears, gas lost from aging stars serves as a substantial source of hot metal-rich interstellar medium (ISM). 
Such enriched hot gas 
radiates mainly in X-rays at a temperature of 10$^{6.4-7.0}$ K via thermal bremsstrahlung with metal line emission. In addition to hot gas, unresolved low mass X-ray binaries (LMXBs) and other stellar components contribute to the diffuse X-ray emission. 

Early-type galaxies with similar optical luminosities ($L_{\rm opt}$) might be expected to have similar X-ray luminosities ($L_{X}$), 
since a comparable amount of stellar mass from their similar stellar populations should release similar amounts of material into the ISM over a given time period.
The surprisingly large dispersion in the $L_{X}/L_{\rm opt}$ relation has been reported since {\sl Einstein Observatory} (e.g. Canizares et al.\ 1987; Fabbiano et al.\ 1992), 
with a variation in X-ray luminosity among early-type galaxies of similar optical luminosities up to two orders of magnitude (O'Sullivan et al.\ 2001). 
Such a discrepancy has been a well-known puzzle in X-ray astronomy for the last three decades.

Another puzzle involves the metallicity of the hot ISM in early-type galaxies. 
The ISM metal abundance traces the history of the star formation and galaxy evolution. 
The ultimate sources of enriched hot gas are red giant winds, planetary nebulae and supernovae ejecta. Given that the stars of early-type
galaxies have measured abundances near solar, and Type Ia supernovae (SNIa) contribute even
higher metallicities, the hot ISM should show an abundance well above solar. 
However, contrary to theoretical derivation, {\sl ASCA} detected an Fe abundance approaching $\sim 0.1$ Z$_{\odot}$ for most elliptical galaxies (Arimoto et al.\ 1997). Such anomalously low abundance values likely resulted from the inability of {\sl ASCA} to spatially resolve the LMXBs components or temperature gradients (the ``Fe bias"\footnote{ It has been shown that Fe abundance is biased low if non-isothermal gas is fit with a single temperature component (Buote 2002).}).

To some extent, the low metallicity issue has been partially resolved at least 
for some X-ray bright galaxies. Current missions such as {\it Chandra},
{\it XMM-Newton} and {\it Suzaku} observe that the metal abundance of the ISM in hot gas-rich galaxies tend to be approximately solar or slightly super-solar (e.g.,
Humphrey \& Buote 2006; Xu et al.\ 2002; Loewenstein \& Davis 2010). However, this is still 
at odds with our classical understanding of the enrichment processes of such
systems. The origin of the metal discrepancy for X-ray bright galaxies has been reviewed in Pipino \& Matteucci (2011, and references therein). One possibility is that a large fraction of SNIa ejecta may condense into dust rather than staying in the X-ray emitting hot phase. 
Fe-enriched gas cools faster than metal-poor gas because of its larger radiative emissivity.
In fact, giant ellipticals may contain up to $10^7 M_{\odot}$ of dust (Temi et al.\ 2004), which makes dust-assisting cooling efficient. 
 However, simulations have revealed that it is unlikely for Fe ejecta to cool and drop out of the hot gas phase (Tang \& Wang 2010). 
Alternatively, a variation in SNIa yields or a large uncertainty in star formation history may also help explain the disagreement. Moreover, Pipino et al.\ (2005) also investigated whether the dilution from freshly accreted cold gas could reduce hot gas metal abundance.

For X-ray faint galaxies (with only small amounts of hot ISM), the metallicity discrepancy is much worse, where very sub-solar
($\sim$10\% solar) abundances have been reported even with current missions
(e.g., NGC~1291, Irwin et al.\ 2002; NGC~4697, 
Sarazin et al.\ 2001; NGC~3585, NGC~4494, NGC~5322, O'Sullivan \& Ponman 2004).
Early-type galaxies of various masses may have intrinsically different stellar Fe abundance 
which may be a factor in driving abundance differences. However, even elliptical galaxies with the same stellar metallicity show up to a factor of ten variation
in ISM metal abundance (Humphrey \& Buote 2006), making such an explanation unlikely.  

The inconsistency with theory and the fact that abundances vary widely among early-type
galaxies of similar optical luminosities
predict a crucial missing (possibly external) factor in the enrichment process.
It has been suggested that galactic winds, SNIa and AGN could drive gaseous outflow (Mulchaey et al.\ 2010; Alatalo et al.\ 2011). There is also observational evidence for gas being removed from galaxies through ram pressure in dense environments (Owen et al.\ 2006; Sun et al.\ 2007). Through such processes, 
enriched gas of early-type galaxies can be transported into their environments, eventually 
enriching intragroup medium (IGrM) or intracluster medium (ICM). It also helps to explain the large scatter in the $L_{X}/L_{\rm opt}$ relation, in the sense that outflows were able to more efficiently remove gas from X-ray faint galaxies.    
However, such a process cannot explain the heavy element deficiency of the remaining gas, since it should work equivalently for both metals and hydrogen. Another explanation associates the dilution by the accretion of the relatively pristine local intergalactic medium.  
However, this explanation faces difficulties when it comes to isolated galaxies not to mention field galaxies tend to have a lower ISM metallicity than cluster galaxies.

The role played by dilution via the accretion of surrounding cold gas in the enrichment processes has not been observationally investigated thoroughly. 
While the classical view of early-type galaxies is that they contain little or
no cold (atomic or molecular) ISM, an increasing number of early-type galaxies
have been found to contain significant amount of 
cold gas in the phase of H~{\sc i} (up to 10$^9$ M$_{\odot}$; Oosterloo et al.\ 2010)
and H$_2$ (up to 10$^9$ M$_{\odot}$; Young et al.\ 2011). 
The fraction of early-type galaxies that are reported to have strong H~{\sc i} absorption is 
around 50\%, which is as significant as the H~{\sc i} detection fraction for star formation galaxies
(Thom et al.\ 2012). Such H~{\sc i} gas is also bound to its host galaxy, with the velocities 
of detected H~{\sc i} below the escape velocity (Thom et al.\ 2012).
Additional studies
show that the H~{\sc i} structures in early-type galaxies can reach out to many
tens of kpc from the stellar body (Helmboldt 2007).
Due to their relatively large orbit, when cold gas is compressed toward the galaxy center, 
their large potential energy difference would be eventually transformed into kinetic energy. Such a process may heat up the cold gas into the X-ray emitting phase.
The overall detection rate of H$_2$ gas is 22\% according to the largest volume-limited CO survey to date (Atlas$^{3D}$ collaboration\footnote{http://www-astro.physics.ox.ac.uk/atlas3d/}). H$_2$ gas is more bound to their host galaxies than H~{\sc i}, and their distributions are more concentrated (Davis et al.\ 2009). Molecular gas could be shock heated to $10^6$ K or higher via relative stellar velocities and interacting with hot gas (Young et al.\ 2011).   
Moreover, field galaxies tend to contain larger H~{\sc i} halo than galaxies in groups and clusters which may have had their H~{\sc i} halo destroyed by stripping or ram pressure (Oosterloo et al.\ 2010), while H$_2$ gas mass of galaxies does not seem to depend on environments (Young et al.\ 2011). 
The origin of such cold gas remains ambiguous. One explanation is that they may be the leftovers from the epoch of galaxy formation, or high angular momentum tidal gas that survived the merging process when early-type galaxies were transformed from spirals and settled into stable orbits around the newly-formed galaxies (Barnes 2002). 
They also have been proposed to originate from recent accretion from filaments, recent mergers and internal star formation (Davis et al.\ 2009). Li et al.\ (2011) provide evidence for the interaction between cold gas and hot gas through SNIa feedback. It is therefore desirable to explore the part played by cold gas in galaxy enrichment processes.

This paper focuses on the metal abundance of X-ray faint early-type galaxies and the metal abundance difference between X-ray faint and the much better studied X-ray bright galaxies. 
As mentioned above, one explanation for this discrepancy is that accreted cold gas may have played a crucial role in the enrichment process of the X-ray hot gas in early-type galaxies by diluting the metal
content of hot gas-poor (hot gas mass $\sim$$1\times10^{7}$ M$_{\odot}$) galaxies
from their original approximately solar value to $\sim$20-30\% solar. 
For galaxies with
such H~{\sc i} structures destroyed (such as cluster galaxies which
may have experienced stripping or ram pressure in a more dense ICM 
environment), the metal abundance of the hot gas should be much closer to solar.
In this scenario X-ray bright galaxies would be relatively unaffected by
dilution with cold gas due to their copious amount of existing hot metal-enriched gas
($\sim$10$^{10}$ M$_{\odot}$), making any cold gas dilution inconsequential.

Many observations of X-ray faint galaxies do not have a sufficiently high X-ray count rate to constrain model parameters well, due to their low X-ray brightness. 
Their unresolved LMXBs and background also contribute a large fraction to the diffuse emission, leading to a poor signal-to-noise ratio for the hot gas component.  
It is natural to suspect that the low measured abundance of X-ray faint early-type galaxies is an outcome of low S/N or some other artificial bias.
Unlike hot gas in the ICM, temperatures of hot gas in the ISM are usually $\approx$ 0.3--1 keV. Fe abundances in this temperature range can only be derived with complicated, incompletely ionized Fe-L lines. This makes the Fe abundance measurement sensitive to instruments, choice of spectral models, background subtraction, etc. We address such artificial factors in this work.

We assume $H_0 = 70$ km s$^{-1}$ Mpc$^{-1}$, $\Omega_{\Lambda}=0.7$, and  $\Omega_M=0.3$.  Throughout this paper, all uncertainties are given at the $90\%$ confidence level unless otherwise stated. We adopt the solar abundance standard of Asplund  (2009), which considered the deviations from local thermodynamic equilibrium.  
 Sample selection is presented in Section 2. Section 3 and 4 are dedicated to observations and data reduction. In Section 5 we report several relationships between galaxy properties. In Section 6 we examined potential biases. Implications of our results are discussed in Section 7. Finally, we summarize our main conclusions in Section 8. 

\section{\bf Sample Selection} 
 \subsection{\bf X-ray observations}

Our ultimate goal is to study the X-ray emitting hot gas properties (e.g., metal abundance) of nearby early-type galaxies as completely as possible. 
We select our sample from a volume-limited sample of 260 nearby early-type galaxies from $ATLAS^{3D}$ plus 61 galaxies studied with {\sl ROSAT} from  Irwin et al.\ (1998). We only considered galaxies that have {\it Chandra} or {\it XMM-Newton} observations with a total exposure time of at least 25 ksec. 
We did not consider any galaxies residing in cluster centers since it is extremely challenging to disentangle the strong ICM emission from relatively weak ISM emission. 
We ultimately selected 32 galaxies with sufficient X-ray counts to have their metal abundance constrained with our model. 
Our sample included galaxies both in groups/clusters and in the field. Classification of galaxy environments is based on Faber (1989). Eight of our galaxies are at group centers (NGC~507, NGC~1399, NGC~4472, NGC~4636, NGC~1332, NGC~4261, NGC~1407, NGC~5846).  
We kept such relatively bright group center galaxies to form the bright end of this relatively continuous sample.
The archived data itself does not form a statistically complete sample.  
Still, our sample of 32 early-type galaxies covers a   
span of $L_X/L_K$ of more than two orders of magnitudes ($0.03$--$3.00\times10^{30}$ ergs s$^{-1}{L_{K,\odot}}^{-1}$),  
containing a sufficient number of X-ray faint, X-ray bright and intermediate-brightness galaxies.
For the majority of the galaxies in this sample we adopted the distance estimation derived from surface brightness fluctuations of Tonry (2001). 
If not available, we used the distance determined from redshift as given in NED\footnote{http://ned.ipac.caltech.edu/} and our assumed cosmology.  
Galaxy properties and observation logs are summarized in Table~1. 

\subsection{Assumed H~{\sc i} and H$_{2}$ gas mass values}
We use the atomic neutral gas (M(H~{\sc i})) mass and molecular gas (M(H$_2$)) mass to 
represent the amount of cold gas mass in each galaxy obtained from the literature. 
Only 24 galaxies in our sample have published masses of atomic and molecular gas as listed in Table~2.  Most of the atomic neutral gas masses were obtained with the Westerbork Synthesis Radio Telescope (WSRT).
 Molecular gas masses were derived with the measurements of $^{12}$CO(1-0) and (2-1) emission lines obtained with IRAM 30-m Telescope. 

\section{\bf Data reduction} 
\smallskip
\subsection{X-ray data}
\subsubsection{Chandra}
We used CIAO4.3 to reduce ACIS-I or ACIS-S data (Table~3). All data were reprocessed from level 1 events, which guarantees the latest and consistent calibrations.  
Only the events with grades 0, 2, 3, 4, and 6 are included. We also removed bad pixels, bad columns, and node boundaries.
We filtered background flares with the light curve filtering script {\tt lc\_clean}.  The effective exposure times are shown in Table~3. 
Bright point sources including nuclei resolved with {\tt wavdetect} were removed. 
In our spectral analysis, each spectrum contains at least 25 counts per energy bin.

\subsubsection{XMM-Newton}
Only data from the European Photon Imaging Camera (EPIC) are reported in this paper (Table~3). Both MOS and PN detectors were processed.
The standard Science Analysis System (SAS 11.0.0) pipeline 
tools were used throughout this analysis. Tasks {\tt emchain} and {\tt epchain} were used to generate calibrated 
event files from raw data. $PATTERN \leq12$ was used to select MOS data sets, while $PATTERN \leq 4$ was used
for PN data sets. The removal of bright pixels and hot columns was done by applying the expression 
(FLAG==0). 
Point sources resolved with {\tt edetect\_chain} and verified by eye were removed. 
The remaining exposure time after filtering for background flares is shown in Table~3. The minimum counts for each energy bin is 25 for MOS and 50 for PN. 

\subsubsection{Regions and Background}
We adopted the effective radii for each galaxy from the Third Reference Catalogue of Bright Galaxies (RC3, de Vaucouleurs et al.\ 1991). 
The extracted aperture for $L_{X, gas}$, $M_{X, gas}$ as well as metal abundance and temperature determinations was chosen to be exactly two effective radii for each galaxy.

Local background, extracted from a region away from the source region on the same ccd chip, is used as background for spectral analysis for both {\it Chandra} and {\it XMM-Newton}. 
The area of the local background was chosen to be at least twice the area of the source region to ensure a sufficient S/N ratio for background subtraction. For X-ray bright galaxies observed with {\it Chandra}, the emission sometimes filled the entire chip due to the relatively small area of the S3 chip. Since X-ray bright galaxies are usually in clusters or at group centers, the adoption of local background enables us to subtract the surrounding ICM or IGrM, assuming the surface brightnesses of ICM/IGrM are uniform on a scale of $\sim$ 20 kpc. 
The variation of background emission is also relatively insignificant for X-ray bright galaxy studies.
For X-ray faint galaxies, the emission usually does not fill the entire chip, which makes it ideal to use local background for spectral studies, unless they are very nearby. One counter example is the X-ray faint galaxy NGC~4697, which is at a distance of only 11 Mpc and has a very extended X-ray emission distribution that extends beyond the S3 chip boundaries. 
For this galaxy, we also tried to use ``stowed background"\footnote{particle background inside the detector observed with ACIS stowed} for the spectral fitting. 
We fit a spectrum extracted from the S1 chip with a stowed background\footnote{http://cxc.harvard.edu/contrib/maxim/stowed} of the same region on the S1 chip to determine the surface brightness of cosmic X-ray and Galactic emission background since the S1 chip is more offset and less contaminated by source emission. Then, we fit the spectrum of NGC~4697 extracted from the S3 chip with a corresponding stowed background by adding scaled X-ray background components obtained with the S1 chip to the fitting.  
The determinations for Fe metal abundance (see model fitting procedure in $\S4.2$) with these two different methods are 
$0.42^{+0.22}_{-0.14}$ Z$_{\odot}$ with $\chi^2_{\nu}=1.01$ (local background) and $0.35^{+0.17}_{-0.13}$ Z$_\odot$ with $\chi^2_{\nu}=1.18$ (stowed background), consistent within the uncertainties.    
Therefore, it is reliable to use local background for galaxies in our sample, even in extreme cases such as NGC~4697.

\subsection{2MASS}
To characterize the optical brightness of each galaxy, we use the K-band luminosity, which is more representative of relatively old stellar populations in early-type galaxies, instead of the historically-used B-band luminosity.   
$L_{K}$ of these galaxies are derived from Two Micron All Sky Survey (2MASS) archived images. The K-band photometry region is the same as that used in the {\it Chandra} and {\it XMM-Newton} analyses. 
Bright nuclear and foreground sources (detected by eye) were excluded and refilled with a local surface brightness component using {\tt dmfilth} in CIAO4.3.  We obtained the counts from the source region after subtracting the local background component. We converted it to the corresponding magnitudes and corrected for Galactic extinction. K-band infrared solar luminosity is assumed to be $L_{K,\odot}=5.67\times10^{31}$ ergs s$^{-1}$ (Mannucci et al.\ 2005). $L_{K}$ of galaxies in this sample are listed in Table~4.

\section{\bf Data Analysis} 
\smallskip

\subsection{Spectral analysis}
\subsubsection{Spectral modelling}
For those galaxies with both {\it Chandra} and {\it XMM-Newton} observations, we conducted joint fits for the measurements of temperature and metal abundance.
We performed spectral analysis with Xspec 12.7.0.
The model we adopted to fit the diffuse emission for each galaxy is
${\tt phabs}*({\tt vapec}+{\tt vapec}+{\tt powerlaw}+{\tt mekal}+{\tt powerlaw})$.   
The absorbing column density ($N_{H}$) was fixed at the Galactic value (Dickey \& Lockman 1990). 
{\tt vapec}+{\tt vapec} represent two temperature components of bremsstrahlung emission from the hot gas, with their elemental abundances tied to each other. We set Mg=Al, Si=S and He=C=N=Ar=Ca=1 (Nagino \& Matsushita 2010; Hayashi et al.\ 2009). We use this two temperature component model to reduce the Fe bias, brought about by multi-temperature gas. The first {\tt Powerlaw} with an index of 1.6 represents the contribution from unresolved LMXBs (Irwin et al.\ 2003). 
In addition to hot gas and unresolved LMXBs, faint stellar X-ray sources such as cataclysmic variables (CVs) and coronally active binaries (ABs) also contribute to $L_{X}$.
Revnivtsev et al.\ (2007, 2008, 2009) calibrated the X-ray emission from such old stellar populations in several extremely gas-poor galaxies.
We estimated such stellar contributions from their $L_{K}$ based on a $L_{X}/L_{K}$ relation averaged over  these gas-poor early-type galaxies given by Revnivtsev et al. (2008): $L_{0.5\text{--}2.0\text{\,keV}}/L_{\text{K}}= 5.9\times10^{27}$\,ergs $s^{-1}$ ${L_{\text{K,}\odot}}^{-1}$.
The {\tt mekal}+{\tt powerlaw} component represents CV/ABs. The temperature of {\tt mekal} is fixed at 0.5 keV, and the index of {\tt powerlaw} is fixed at 1.9 (Revnivtsev et al.\ 2008). In our spectral analysis, we fixed such CV/ABs components at the estimated flux based on the $L_{K}$ of each galaxy. The ratio of the fluxes of the {\tt mekal} component and the {\tt powerlaw} was set to 2.03 (Revnivtsev et al.\ 2008).

We tested the effects of our spectral model and assumptions on NGC~4459, which is the X-ray faintest galaxy in this sample. The two temperature thermal components model gives a Fe abundance of $0.22^{+0.09}_{-0.06}$ Z$_{\odot}$. After setting Al $\neq$ Mg and S $\neq$ Si, we obtained a Fe abundance of  $0.23^{+0.07}_{-0.07}$. The calibration of the X-ray emission of CV/ABs is not well determined, and varies between $L_X/L_K=4.1-6.9 \times 10^{27}$ for M32, N3379, and M31 (Revnivtsev et al.\ 2007; 2008). We varied the CV/ABs component by 100\% by completely ignoring it and by doubling the contribution of such component, which gives a Fe abundance of $0.23^{+0.07}_{-0.06}$ Z$_{\odot}$ and $0.22^{+0.07}_{-0.07}$ Z$_{\odot}$, respectively, indicating that uncertainties in the CV/AB normalization are not relevant to our metal abundance determinations.

\subsubsection{Joint fitting Chandra and XMM}
In order to minimize statistical uncertainty,
if available, {\it Chandra} and {\it XMM-Newton} observations were jointly fit for each galaxy, with all normalizations varied independently, but only the flux and normalization of {\it Chandra} are used to determine $L_X$ and $M_{X, gas}$ \footnote{for NGC~5982 {\it XMM-Newton} results are used to determine $L_X$ and $M_{X, gas}$, since {\it Chandra} data are not available.}, since we left all normalizations of each data set untied.  We use 0.5--8.0 keV for ACIS-I, ACIS-S and PN, 0.3--8.0 keV for MOS to fit the spectra. 
It is important to justify that the cross calibrations are sufficient to provide reliable abundances. 
We examined all galaxies in our sample by determining their hot gas Fe abundance within two effective radii separately with {\it Chandra} and {\it XMM-EPIC}. Among them, 15 galaxies contain sufficient data to determine temperatures and metallicities separately from either data set. All of these 15 galaxies show consistency within the uncertainties between {\it Chandra} and {\it XMM-EPIC} results. Overall, this cross check gives us confidence in our joint fitting strategy. 

 \subsection{Determination of $L_{X, \rm gas}$ and $M_{X, \rm gas}$}
$L_{X, \rm gas}$ estimated in this paper is contributed only by hot gas from $0.1-2.0$ keV, excluding X-ray emission contributed by CV/ABs, and unresolved LMXBs, which have been removed spectrally.  
Assuming a spherical distribution of hot gas,  we obtained their volume from the size of extraction region.
Based on the sum of the best fit normalizations of the thermal emission model {\tt vapec}+{\tt vapec}, we derived the average hot gas density. 
With hot gas density and volume, we obtained hot gas mass ($M_{X, \rm gas}$) for galaxies in this sample.
We assume that the hot gas density is a single value in the given volume for each galaxy. To test how a density gradient affect the result, we divide NGC~720 into 10 spatial bins within two effective radii. We analyzed each bin separately and obtained a sum of gas masses of the 10 bins that is within $10\%$ of the gas mass obtained by analyzing a single integrated bin of a size of two effective radii.

\section{\bf Results}
\smallskip
In Figure~1, we plot $L_{X,\rm gas}$ for each galaxy as a function of $L_{K}$ 
, which shows the scatter of $L_{X,\rm gas}/L_{K}$ is more than a factor of 50. This factor is in line with but somewhat smaller than previously found (e.g. Boroson et al.\ 2011). Since our work aims at studying hot gas Fe abundance only in galaxies with sufficient X-ray counts, we eliminated galaxies that are extremely X-ray faint, leading to a smaller scatter in 
the $L_{X,\rm gas}$--$L_{K}$ relation.
The span of $L_{X,\rm gas}$ is from $7\times10^{38}$ ergs s$^{-1}$ to $1.7\times10^{42}$ ergs s$^{-1}$; the span of $L_{K}$ is from $1.5\times10^{10}L_{\odot}$ to $5.6\times10^{11}L_{\odot}$ (Table~4). Galaxies that reside at group centers are among the brightest. Galaxies in the field are on average fainter than galaxies in groups and clusters.

Hot gas Fe abundance ranges from 0.22 Z$_{\odot}$ (NGC~4459) to 1.9 Z$_{\odot}$ (NGC~1407) (Table~5; Figure~2).
 The Fe abundance generally increases with $L_{X, gas}/L_{K}$ for most galaxies (Figure~2), which gives 
a Spearman correlation coefficient of $\rho=0.449$ with a null hypothesis probability of $1.24\%$ 
for the $L_{X, gas}/L_{K}$--Fe relation using the ASURV software package\footnote{http://astrostatistics.psu.edu/statcodes/asurv}.
The slope becomes flatter for X-ray bright galaxies such as those at group centers. 
Excluding group centers, we obtained a Spearman correlation coefficient of $\rho=-0.673$ with a null hypothesis probability of $0.12\%$. 
We also  break galaxies in our sample into various environments as galaxies in the field, galaxies in groups and clusters but not at centers, and galaxies that reside at group centers, although only two galaxies in our sample are in field. 
A gradient in luminosity, temperature and hot gas Fe abundance can be seen as galaxies in denser environments are brighter, hotter and Fe richer.

Only 22 galaxies in this sample have published Lick/IDS index measurements (listed in Table~1) from which we obtain the stellar metallicity [Fe/H] (Thomas et al.\ 2011; Maraston et al.\ 2011; Thomas et al.\ 2010; Johansson et al.\ 2011). The relation between the hot gas Fe and stellar metallicity is indeed very random (Figure~3 (a)), although NGC~1407, the galaxy with the highest hot gas Fe, does have the highest stellar metallicity. 
For galaxies of similar stellar abundance, the scatter of their ISM Fe is up to a factor of five.
Therefore, the discrepancy in ISM abundance among galaxies is unlikely to be a result of the variation in stellar metallicity. 
We also tried to associate hot gas Fe abundance with stellar ages which we obtained from the literature (listed in Table~1). There is virtually no correlation between these two variables as shown in Figure~3 (b). Therefore, the variation of hot gas Fe abundance is unlikely to be caused by differing galaxy ages either.

To test the cold gas dilution scenario, we studied a sub sample of 24 galaxies with cold gas data available. We related the hot gas Fe abundance to the ratio of their atomic gas mass M(H~{\sc i}) to hot gas mass $M_{X,\rm gas}$ and to the ratio of  molecular gas mass $M(H_2)$ to hot gas mass $M_{X,\rm gas}$, respectively, (M(H~{\sc i})$/$$M_{X,\rm gas}$ and M($H_2$)$/$$M_{X,\rm gas}$). 
Such a ratio is larger for X-ray faint galaxies at a given cold gas mass. 
We find that there is virtually no correlation between Fe and M(H~{\sc i})$/$$M_{X,\rm gas}$ as shown in Figure~4(a); using {\sl ASURV} which takes into account upper limit of variables, we obtained 
a Spearman correlation coefficient of $\rho=-0.089$ with a null hypothesis probability of $67.1\%$, illustrating the limited effect of atomic gas on the hot gas metal content. 
In contrast, we find that Fe abundance generally decreases with $M(H_2)/M_{X,\rm gas}$ as shown in Figure~4(b); we obtained 
a Spearman correlation coefficient of $\rho=-0.459$ with a null hypothesis probability of $2.76\%$. 
We summarized the correlations for each relation in Table~6. 
\section{Possible abundance determination biases}

Previously determined extremely low Fe abundances for NGC~1291 (Irwin et al.\ 2002) and NGC~4697 (Sarazin et al.\ 2001) are not confirmed because in our analysis we free individual elements, take account of CV/ABs, allow multi-temperature components and adopt a revised abundance table.  Only by eliminating such improvements, can we recover the $\sim10\%$ Fe abundances for these galaxies found by previous studies. Here, we discuss some additional potential biases to abundance determinations.

\subsection{``Frankenstein" test}

One potential bias that may affect abundance determination may originate from the different ratios of X-ray emitting components for X-ray faint and X-ray bright galaxies. The fraction of X-ray emission produced by hot gas is smaller for X-ray faint galaxies than for X-ray bright galaxies. For example, the ratio of hot gas to unresolved LMXBs to CV/ABs to background of a typical X-ray bright galaxy is 83:5:2:10  while that of a typical X-ray faint galaxy is 20:20:10:50.
To explore this effect, we built a spectrum of a typical X-ray faint galaxy out of the X-ray components of a typical X-ray bright galaxy by separating and rearranging the contributions from the hot gas, LMXBs, CV/ABs, and background: a so-called ``Frankenstein" galaxy. 

We chose NGC~720 as a representative X-ray bright galaxy. The X-ray emission from this galaxy is partitioned as follows: 5.7\% unresolved LMXB,  62.5\% gas, 1.8\% CV/AB and 30\% background. NGC~720, as a galaxy in the field, is not influenced by any environmental effects such as ram pressure, tidal disruption, or galaxy harassment. The morphology of NGC~720 also appears very symmetric. 
The temperature radial profile of this galaxy shows little gradience
(Humphrey et al.\ 2011), which minimizes any biasing in the measurement of  Fe abundance due to multi-temperature components. All these factors assure a relatively robust metal abundance measurement. 

There are five main {\it Chandra} ACIS-S observations of NGC~720 with comparable exposure times. We combined the spectra of resolved point sources\footnote{the brightest point sources were not included. Such point sources can definitely be resolved by {\it Chandra} for all galaxies in our sample, and they may have a spectral shape different from the unresolved fainter LMXBs (Irwin et al.\ 2003).} from all five observations for the LMXBs spectrum. 
The spectrum of the hot gas was extracted from only one observation (the extracted region is the same as $\S4.1.2$). The background component was extracted from various source-free off-center regions of three observations. 
We then simulated the spectrum of CV/ABs ({\tt mekal}+{\tt powerlaw}). Combining these spectra produced a typical X-ray faint galaxy spectrum that contains 23\% hot gas, 20\% LMXBs, 10\% CV/ABs and 47\% background. We fit this Frankenstein spectrum with a separate local background spectrum 
 and compared it with the results of the original X-ray bright galaxy NGC~720. The best fit Fe for this Frankenstein galaxy is $0.79^{+1.31}_{-0.31}$ Z$_{\odot}$, while that of the original X-ray bright galaxy NGC~720 is $0.91^{+0.22}_{-0.16}$ Z$_{\odot}$, consistent within the uncertainties. Therefore, the ratio of spectral components is unlikely to cause the measured low abundance of X-ray faint galaxies.

\subsection{Complex temperature structure}

When fitting intrinsically multi-temperature systems with a single temperature model, the measured Fe abundance is systematically low, and is sometimes referred to as the Fe bias (Buote 2002). To address how significant this effect is, 
we tried to fit the hot gas component with just a single {\tt vapec} model for each galaxy. 
We compared the Fe abundance as measured with a single temperature model (1T) to the Fe abundance measured with a two temperature model (2T). The measured Fe abundances are plotted in Figure~5. 
The Fe abundance measured with 2T is on average 20\% greater than with 1T. 
In Figure~6 (a), we show the ratio of Fe abundance determined with 2T model and that with 1T model as a function of $L_{X, \rm gas}/L_K$. ``Fe bias" does bias low the measured hot gas Fe abundance but seems equally effective for both X-ray faint and X-ray bright galaxies, therefore, unlikely to cause the Fe abundance discrepancy between X-ray faint and X-ray bright galaxies.

We also simulated a spectrum with a model of ${\tt phabs}*({\tt cevmkl}+{\tt powerlaw}+{\tt mekal}+{\tt powerlaw})$, where the Xspec model {\tt cevmkl} represents  a continuously varying multi-temperature hot gas emission. The Fe abundance of {\tt cevmkl} was set to 1.0 Z$_{\odot}$. The simulated spectrum was built to have the same counts rate, exposure times, CV/ABs and LMXBs as NGC~720. We fit this simulated spectrum with the 2T model: 
${\tt phabs}*({\tt vapec}+{\tt vapec}+{\tt powerlaw}+{\tt mekal}+{\tt powerlaw})$ as described earlier. We still obtained a Fe abundance of {\tt vapec}+{\tt vapec} to be around 1.0 Z$_{\odot}$, even when we vary the power-law index for the temperature distribution and the maximum temperature in {\tt cevmkl} by at least 50\%. This test demonstrates that the 2T model we applied is sufficient to describe continuous varying multi-temperature structures.  

However,  the hot gas Fe abundance may indeed be underestimated due to more complex temperature and abundance structures, such that even a 2T model might be insufficient to reflect the real value of the Fe abundance. 
It is impractical to fit galaxies in our sample with 2T/2Fe model with the limited S/N of out data set.
Therefore, we simulated a spectrum based on NGC~720
with a 2T model. We let each {\tt vapec} component have different temperatures and Fe abundances but the same flux (2T/2Fe model:
T${_1}$=0.2, Fe${_1}$=1.4 Z$_{\odot}$ and T$_2$=0.6, Fe$_2$=0.4 Z$_{\odot}$; higher metallicity gas is supposed to cool faster). We fit this spectrum with a 2T model but tied the Fe abundance of each {\tt vapec} to each other. We obtained a Fe abundance of only 0.5 Z$_{\odot}$, smaller than the emission weighted average abundance of the two {\tt vapec} but still between the bookend values. The measured Fe abundance is further reduced by 25\% if we fit this spectrum with a 1T model.
We did the same test based for NGC~4697 (2T/2Fe model: T${_1}$=0.2, Fe${_1}$=0.8 Z$_{\odot}$ and T$_2$=0.4, Fe$_2$=0.2 Z$_{\odot}$), and obtained a best fit Fe abundance of 0.22 Z$_{\odot}$ with 2T model for this simulated spectrum. The Fe abundance was underestimated to the same degree for NGC~720 and NGC~4697.   
In our work, we focus on the relative difference of Fe abundance between X-ray faint and X-ray bright galaxies instead of their absolute values. Such a uniform underestimation for Fe abundance should not seriously affect our conclusions. Exploring a larger Fe/T parameter space modeling and a more realistic simulation are needed in future work to address this point more fully.   

\medskip

\section{\bf Discussion}

In this paper we have demonstrated a discrepancy in the hot gas metal abundance between X-ray faint and X-ray bright galaxies. 
As shown in Figure~2, it is evident that abundance increases with $L_{X, \rm gas}/L_{K}$ for X-ray faint galaxies but less so for X-ray brightest galaxies. This indicates that X-ray faint galaxies are more affected by whatever the origin of this variation is. 
In this work, we found that there is essentially no correlation between hot gas Fe abundance and the ratio of atomic gas mass to hot gas mass (M(H~{\sc i})$/$M$_{X,\rm gas}$). Hot gas Fe abundance does not appear to be related to M(H~{\sc i}) either as shown in Table~6. Thus, this would seem to rule out the scenario that the accretion of pristine atomic gas could affect the metal content of hot gas of early-type galaxies. Atomic gas is usually distributed in large orbits and in some cases they are in a shape of a ring  surrounding the galaxy (e.g. NGC~1291). The accretion of such outskirt gas may not be efficient. Moreover, such accretion may be suppressed during the process of gaseous outflows.

In contrast, we found a significant anti-correlation between the hot gas Fe abundance as measured in X-ray and the ratio of molecular gas mass to hot gas mass (M$(H_{2}$)$/$M$_{X,\rm gas}$); galaxies with a larger molecular gas fraction tend to have lower Fe abundance while galaxies with a smaller cold gas fraction show a higher Fe abundance (Figure~4). The Spearman correlation coefficient is $\rho=-0.459$ with a null hypothesis probability of $2.76\%$. Hot gas Fe is also correlated with M($H_2$) as shown in Table~6.  Unlike atomic gas, molecular gas is usually located in the inner regions of galaxies which makes it easier for it to interact with hot gas. This result seemingly suggests a scenario where the
molecular gas has been shock heated to the X-ray emitting phase by relative velocities between stellar wind and ambient hot gas. 
However, unlike atomic gas, it is unjustified to assume that molecular gas is pristine. In fact, molecular gas is usually associated with new star formation, which makes its metallicity no less than that of hot gas, putting this dilution scenario in doubt.

 To search for an alternative explanation, we first examined the reliability of the molecular gas mass estimation.  M($H_2$) adopted in this paper were derived from CO emission lines (e.g, Young et al.\ 2011). Those authors assume a constant conversion between M($H_2$) and CO emission. In fact, M($H_2$)-F$_{CO}$ is a function of stellar metallicity as lower metallicity systems have a higher M($H_2$)$/$F$_{CO}$ ratio (Genzel et al.\ 2012). However, as illustrated in $\S5$,
stellar metallicity and hot gas metallicity are not directly related. Therefore, this bias should not be responsible for this apparent anti-correlation between hot gas Fe and M$(H_{2}$)$/$M$_{X,\rm gas}$. Moreover, even if higher stellar metallicity galaxies tend to have higher hot gas Fe abundance, the true anti-correlation between hot gas Fe -- M($H_2$)$/$M($_{X,\rm gas}$) would have been even stronger because assuming a constant conversion for all galaxies could only underestimate the M($H_2$) for lower metallicity systems.

 Second, this apparent anti-correlation may be a result of complex temperature structures in a sense that a larger molecular gas fraction may lead to a more non-uniform temperature distribution of hot gas through cooling and therefore bias low the measurement of hot gas Fe abundance. This scenario predicts that the ratio of Fe abundance determined with a 2T model to that with a 1T model should depend on molecular gas fraction. However, as shown in Figure~6(b), these two factors are not related. Therefore, we do not think complex temperature structure drives this hot gas Fe -- M($H_2$)$/$M($_{X,\rm gas}$) anti-correlation. Still, it could be a result of complex abundance structure as proposed in $\S6.2$. The interaction with molecular gas may cause multi-phase metal abundance distributions which would bias low the measured hot gas Fe abundance. The test of this scenario is beyond the scope of this paper. We expect more simulation works in the future to cast light on this issue.

Third, this apparent anti-correlation may stem from a third factor that is linked to both molecular gas and hot gas Fe abundance. 
Whatever this third factor is, the question why there is a discrepancy of the hot gas Fe metallicity between X-ray faint and X-ray bright galaxies remains to be answered. 
Using 3D simulations, Tang \& Wang (2010) found that SN ejecta has a tendency to move outward substantially faster than the ambient medium via buoyancy force and effectively reduces the average Fe abundance of hot gas.    
If this process is plausible, we can conclude that systems with smaller angular momentum and larger potential wells are more likely to retain their Fe abundance since it is easier for them to resist such buoyancy force.  X-ray bright galaxies are usually massive galaxies which also tend to be slow rotating galaxies (Emsellem et al.\ 2011). Consequently, X-ray bright galaxies would have a larger Fe abundance than X-ray faint galaxies in this scenario. 
It has been speculated that slow rotating galaxies as well as massive galaxies contain less molecular gas (Young et al.\ 2011). Hence, X-ray bright galaxies tend to lack molecular gas. As a result, molecular gas and hot gas Fe abundance appear to be anti-correlated. Yet, it is unclear how dynamic mass, angular momentum and molecular gas content of galaxies are physically related.

Finally, we still try to explore the possibility that  molecular gas may have diluted the metal content of hot gas. In addition to internal stellar mass loss, molecular gas in early-type galaxies have also been proposed to originate externally such as accretion through filaments and mergers with late-type galaxies. Davis et al.\ (2009) show that quite a few galaxies have their molecular gas kinematically misaligned with respect to the stars, suggesting external origin.  Combes et al.\ (2007) inferred that CO-rich galaxies may be more metal and $\alpha$-element poor owing to a slow star formation fuelled by relatively pristine gas. 
Davis et al.\ (2009) also found that the molecular and atomic gas are always kinematically aligned. Therefore, it is possible that molecular gas in early-type galaxies may originate from the condensation of surrounding atomic gas. In that case, the molecular gas might be relatively less contaminated and potentially dilute the hot gas metal content.

There are many other factors that may contribute to the discrepancy between X-ray faint and X-ray bright galaxies as well as the apparent anti-correlation between hot gas Fe abundance and molecular gas content. In fact, early-type galaxies may be quite heterogeneous with various assembly and enrichment histories. Both a case-to-case study and a larger and more complete sample are required to address such issues in the future. Nevertheless, various explanations are not exclusive to each other. A combined scenario may eventually solve the discrepancy.

\section{\bf Summary}
   We studied a sample of 32 early-type galaxies with quality {\it Chandra} and {\it XMM-Newton} data covering a large span of X-ray luminosities. We derive a number of their properties including $L_{X}$, $L_{K}$, temperature, and ISM Fe metallicity. We attempt to relate these properties to their stellar metallicity, stellar age, and cold gas masses, 
to investigate the causes of the low metallicity of hot gas in X-ray faint galaxies and the metal abundance difference between X-ray faint and X-ray bright galaxies. 
We summarize our main results as follows: 
 
$\bullet$ Galaxies with similar $L_{K}$ are observed to have $L_{X, \rm gas}$ that vary more than a factor of 50 generally in agreement with previous studies.    

$\bullet$ Hot gas Fe abundances of early-type galaxies are mostly lower than solar abundance with the brighter/hotter galaxies having a higher Fe abundance than fainter/cooler galaxies. 
This variation does not originate from the variations in stellar metallicities or stellar ages.

$\bullet$ Extremely low Fe abundance ($\sim$ 0.1 Z$_{\odot}$) of early-type galaxies found by previous studies was not confirmed in this work. Nearly all galaxies in our sample have a Fe abundance of at least 0.3 Z${_\odot}$ thanks to the adoptions of corrected models and an updated abundance table. 

$\bullet$ low X-ray count rate and the non-gaseous components of the X-ray emission do not substantially bias measurement of hot gas Fe abundance. 

$\bullet$ Hot gas Fe abundances of early-type galaxies and their atomic gas mass M(H~{\sc i}) are not related. This puts a strong upper limit on the role played by the accretion of atomic gas mass in the metal content of hot gas. 

$\bullet$ Early-type galaxies that have a larger ratio of cold gas mass M(H$_2$) to hot gas mass $M_{X,\rm gas}$ tend to have a lower Fe abundance. However, it is not clear what is the cause of this anti-correlation.

Deeper observations of X-ray as well as radio observations of more early-type galaxies, especially for galaxies in the field, are needed to further test the role played by neutral gas more thoroughly.

\section{Acknowledgments}

We are grateful to Daniel Thomas and Jonas Johansson for calculating stellar metallicities. We thank Dong-Woo Kim and Raymond White for useful discussions and suggestions. 
We thank Evan Million, Ka-Wah Wong, Milhoko Yukita and Zhiyuan Li for reading an early draft and helpful comments.

 \begin{deluxetable}{llllll}
\tablecolumns{6} 
\tablewidth{0pc}
\tabletypesize{\scriptsize}
\tablecaption{Sample properties}
\tablehead{
\colhead{Name}&\colhead{Morphology}&\colhead{$D$}&
\colhead{ $N_{\rm H}$ } &\colhead{ $r_{e}$}&\colhead{Environment$^{\ast}$}\\
\colhead{}&\colhead{} & \colhead{(Mpc)} & \colhead{($10^{20}$ cm$^{-2}$)} & 
\colhead{(arcmin)} & \colhead{}}
\startdata 

IC~1459 & E3-4 & 29.2 & 1.1 & 0.57 &1\\
NGC~507 & SA0 & 70.0 & 5.3 & 0.89 & 2\\
NGC~720 & E5 & 27.7 & 1.5 & 0.60 & 0\\
NGC~1291 & SB0 & 10.1 &  2.1 & 0.90 & 0\\
NGC~1316 & SAB0 & 20.9 & 1.9 & 1.34 &1\\
NGC~1332 & S0 & 22.9 & 2.2 & 0.46 & 2\\
NGC~1380 & SA0 & 17.6 & 1.4 & 0.66 & 1\\
NGC~1395 & E2 & 24.8 & 2.0 & 0.81 & 1\\
NGC~1399 & E1 & 20.0 & 1.5 & 0.67 & 2\\
NGC~1404 & E1 & 21.0 & 1.5 & 0.40 & 1\\
NGC~1407 & E0 & 28.8 & 5.4 & 1.17 &2\\
NGC~3607 & SA0 & 22.8  & 1.4 & 0.72 & 1\\
NGC~3608 & E2 &  22.9  & 1.4 & 1.12&1\\ 
NGC~3923 & E4-5 & 22.9 & 6.2 & 0.83 & 1\\
NGC~4125 & E6 & 23.9 & 1.8 & 0.97 & 1\\
NGC~4261 & E2-3 & 20.0 & 1.8 & 0.60 & 2 \\
NGC~4278 & E1-2 & 16.1 & 1.8 & 0.57 & 1\\
NGC~4365 & E3 & 20.4 & 1.7 & 0.80 & 1\\
NGC~4374 & E1-2 & 18.5 & 2.6 & 0.85 & 1\\
NGC~4382 & SA0 & 18.5 & 2.5 & 0.91 & 1\\
NGC~4406 & E3 & 17.1 & 2.7 & 1.73 & 1\\
NGC~4459 & SA0 & 7.8  & 4.7 & 0.59 & 1\\
NGC~4472 & E2 & 16.3 & 1.5 & 1.73 &2\\
NGC~4477 & SB0 & 16.5 & 2.6 & 0.63  &1\\
NGC~4526 & SAB0 & 16.9 & 1.5 & 0.74 &1\\
NGC~4552 & E0-1 & 15.3 & 2.6 & 0.49 &1\\
NGC~4636 & E0-1 & 14.7 & 1.9 & 1.47 &2\\
NGC~4649 & E2 & 16.8 & 2.2 & 1.15 &1\\
NGC~4697 & E6 & 11.8 & 2.1 & 1.20  &1\\
NGC~5846 & E0-1 & 24.9 & 4.3 & 1.04  &2\\
NGC~5982  & E3 & 42.0 & 1.5 & 0.80 & 1\\
NGC~7619 & E & 53.0 & 5.0 & 0.62 &1\\

\enddata
\tablecomments{$^{\ast}$ 0: Galaxies in the field. 1: Galaxies in groups and clusters but not at centers. 2: Galaxies at group center  (Faber et al.\ 1989)}
\end{deluxetable}

\begin{deluxetable}{lllllll}
\tabletypesize{\scriptsize}
\tablewidth{0pc}
\tablecaption{H~{\sc i}, H$_{2}$, stellar metallicity values and ages}
\tablehead{
\colhead{Name} &\colhead{ $M (HI)$} &\colhead{ $M (H_2)$}& \colhead{Mgb} & \colhead{Fe5270 }& \colhead{Fe5335}& \colhead{Age}\\
\colhead{}& \colhead{($10^{8}M_{\odot}$)} &\colhead{ ($10^{8}M_{\odot}$)}&\colhead{} &\colhead{} &\colhead{} &\colhead{(Gyr)}}
\startdata
IC~1459 &$2.5^{a}$ & $<5.57^k$& $5.60 \pm0.02^f$& $3.36 \pm0.02^f$& $2.97\pm0.02^f$&${3.5^{+1.7}_{-0.4}}^a$ \\
NGC~507 & & &  $4.52 \pm 0.11^n$ & $2.95 \pm 0.12^n$ &$2.6 \pm 0.15^n$&${9.2^{+0.2}_{-0.2}}^o$ \\
NGC~720 &$< 0.59^{b}$&$< 0.24^{b}$&$5.17 \pm 0.11^n$&$2.94 \pm 0.12^n$&$2.80 \pm 0.14^n$&${3.9^{+2.1}_{-2.1}}^a$\\
NGC~1291 & $8.1^j$ & $0.22^j$ & & &&${7.1^{+1.63}_{-1.32}}^q$\\
NGC~1316 & $< 1.60^{k}$&$5.23^{k}$ & $4.07 \pm 0.03$$^f$& $3.13 \pm 0.02^f$ & $2.64 \pm 0.03^f$&$3.2^p$\\
NGC~1332 & & & & & &${4.1^{+5.4}_{-0.9}}^r$ \\
NGC~1380 & & & & &  &${4.4^{+0.7}_{-0.7}}^a$\\
NGC~1395 && &  $4.95 \pm 0.04$$^f$& $2.98 \pm 0.03$$^f$& $2.72 \pm 0.04$$^f$&${6.0^{+1.6}_{-0.9}}^a$ \\
NGC~1399 && &  $5.74 \pm 0.11$$^f$ & $3.15 \pm 0.08$$^f$&$2.85 \pm 0.09$$^f$&${11.5^{+1.1}_{-2.1}}^r$\\
NGC~1404 & & &  $4.65 \pm 0.10 $$^f$& $2.92 \pm 0.07$$^f$&$2.59 \pm 0.08$$^f$&${12.0^{+2.0}_{-3.0}}^a$\\
NGC~1407 & $< 1.52^{b}$ & $< 0.34^{b}$  & $4.94 \pm 0.09^l$ & $3.65 \pm 0.14^l$&$3.37 \pm 0.18^l$&${6.6^{+1.9}_{-1.5}}^a$\\ 
NGC~3607 & $< 0.83^{e}$ & $2.63^{g}$&$4.43 \pm 0.08^l$&$3.37 \pm 0.03^l$&$3.19 \pm 0.10^l$&${3.1^{+0.5}_{-0.5}}^s$\\
NGC~3608 & $ 0.03^{e}$ & $< 0.38^{g}$&&&&${10.7^{+0.5}_{-0.9}}^v$\\
NGC~3923 & & & $4.54 \pm 0.08^l$ &$3.33 \pm 0.10^l$&$0.54 \pm 0.08^l$&${2.8^{+0.3}_{-0.1}}^r$\\
NGC~4125 & $< 0.37^b$ &$<0.34^b$& $4.37\pm0.05^m$&$3.02 \pm 0.05^m$&$3.06 \pm 0.06^m$&${13^{+0.5}_{-0.5}}^r$\\
NGC~4261 &$< 2.95^{h}$&$< 0.49^{h}$&$5.11 \pm 0.04^n$&$3.14 \pm 0.05^n$&$2.88 \pm 0.08^n$&${12.6^{+0}_{-0.6}}^a$\\
NGC~4278 &$6.9^{i}$&$< 0.08^{b}$&$5.24 \pm 0.07^m$&$2.79 \pm 0.07^m$&$2.83 \pm 0.08^m$&${14.1^{+1.4}_{-1.2}}^v$ \\
NGC~4365 &$< 0.4^{a}$ &$<0.42^{g}$& $4.74 \pm 0.07^n$&$3.48 \pm 0.03^n$&$3.13 \pm 0.14^n$ &$5.9^x$\\
NGC~4374 & $< 0.4^{e}$&$< 0.16^{g}$& $4.78 \pm 0.03^n$&$2.94 \pm 0.04^n$&$2.69 \pm 0.04^n$&${12.7^{+2.0}_{-2.0}}^n$\\
NGC~4382 & $< 0.09^{e}$ & $< 0.25^{g}$  & & && ${5.4^{+0.5}_{-0.5}}^v$\\
NGC~4406 & $1^{e}$&$< 0.25^{g}$ &$4.95 \pm 0.07^m$ & $3.04 \pm 0.07^m$ &$2.91 \pm 0.09^m$ &$11^u$\\
NGC~4459 & $< 0.08^{e}$ & $1.73^{g}$ & & & &${7.1^{+0.7}_{-0.6}}^k$\\
NGC~4472 & $< 0.83^{h}$ & $0.006^{h}$ & $4.85 \pm 0.06^n$ &$2.97 \pm 0.07^n$ &$2.84 \pm 0.08^n$& ${7.6^{+1.7}_{-1.6}}^a$\\
NGC~4477 & $< 0.09^{e}$ & $0.35^{g}$ & & & &$9.6^y$\\
NGC~4526 & $21.37^{h}$&$1.1^{h}$ & & & & $1.6^z$\\
NGC~4552 & $< 0.07^{e} $&$< 0.19^{g}$&$5.15 \pm 0.03^n$&$3.02 \pm 0.04^n$&$2.95 \pm 0.04^n$ & ${12^{+1.0}_{-1.0}}^r$\\
NGC~4636 & $10^{h}$& $0.01^{h}$&& & &${7.6^{+3.0}_{-3.0}}^a$\\
NGC~4649 & $< 0.15^{e}$ &$< 0.28^{g}$&$5.33 \pm 0.04^n$&$3.01 \pm 0.04^n$&$3.01 \pm 0.05^n$& ${13^{+1.0}_{-1.0}}^r$\\
NGC~4697 & $< 0.42^{b}$ & $< 0.30^{b}$ & $4.08 \pm 0.05^n$ & $2.97 \pm 0.06^n$ &$2.57 \pm 0.06^n$& ${7.2^{+1.7}_{-1.5}}^a$\\
NGC~5846 & $3.5^{a}$&$< 0.60^{g}$&$4.93 \pm 0.05^n$&$2.95 \pm 0.06^n$&$2.77 \pm 0.06^n$& ${12^{+0.6}_{-1.6}}^a$\\
NGC~5982 & $0.34^{i}$ & $< 0.57^{w}$&&&&${10.7^{+0.5}_{-0.7}}^{v}$\\
NGC~7619 &&&$5.06 \pm 0.04^n$&$3.03 \pm 0.05^n$&$3.08 \pm 0.06^n$ &$12.9^u$\\
 \enddata 
\tablerefs{ a. Serra \& Oosterloo (2010), b. Welch (2010), c. Haynes (1988), d. Taniguchi (1994), e. Serra (2011), f. Beuing (2002), g. Young (2011), h. http://goldmine.mib.infn.it/, i. Oosterloo (2010),
j. Li (2011),
k. Lees (1991), l. Emsellem (2011), l. Denicolo (2005), m. Serra (2008), n. Trager (2000),
o. Li (2006), p. Terlevich (2002), q. Noll (2009), r.  Humphrey (2008),
s. Annibali (2001), t. Sil'chenko (2006), u. Zhang (2008), v. Kuntschner (2010), w. Shapiro (2009), x. Howell (2005), y. McDermid (2006), z. Gallagher (2008)}\\
 \end{deluxetable}

\begin{deluxetable}{llllll}
\tablewidth{0pc}
\tabletypesize{\scriptsize}
\tablecaption{Observational log}
\tablehead{
\colhead{Name} &  \colhead{Chandra}&\multicolumn{2}{c}{Exposure time (ks)}&\colhead{XMM}&\colhead{Exposure time (ks)} \\
\colhead{}&\colhead{Obs-ID}&\colhead{ACIS-S}&\colhead{ACIS-I}&\colhead{Obs-ID}&\colhead{MOS1, MOS2, PN}}
\startdata
IC~1459 & 2196 & 54 & & 0135980201 & 29, 29, 24\\
NGC~507 & 317, 2882 & 15 & 43 & 0080540101 & 34, 34, 30 \\
NGC~720 & 492, 7062  &31, 23 & & 0112300101 & 27, 28, 19\\
&7372, 8448 & 49, 8 && 0602010101 & 84, 84, 69 \\
& 8449 & 19 & & & \\
 & 8198, 8464 & 47, 47 & & & \\
 & 8465 & 45 & & & \\
NGC~1291 & 795, 2059 & 34, 22 & & 0201690201 & 19, 19, 18\\
& 11272, 11273 & 68, 49 & & & \\
NGC~1316 & 2022 & 27 & & 0302780101& 87, 85, 49\\
&&&& 0502070201 & 49, 48, 29\\
NGC~1332 & 2915, 4372 & 16, 43 & & 0304190101& 59, 59, 50\\
NGC~1380 & 9526&39 & & 0210480101&46, 46, 40\\
NGC~1387 & 4188 & & 42 & & \\
NGC~1395 &799 & & 16& 0305930101 & 57, 57, 37 \\
NGC~1399 & 319, 9530 & 54, 54&  & 0040060101 & 114, 111, 89 \\
NGC~1404 & 2942, 4147 & 28 & 45 & 0304940101 & 30, 16, 22 \\
NGC~1407 & 791 & 43 & & 0404750101 & 41, 39, 32 \\
& & & & 0110930201 & 9, 8, 5 \\
NGC~3607/3608 & 2038 & & 27  &0693300101&38, 38, 26 \\
NGC~3923 & 1563, 9507 & 17, 78 & & 0027340101 & 38, 38, 32 \\
&&&& 0602010301 & 99, 99, 76\\
NGC~4125 & 2071 & 62 & & & \\
NGC~4261 & 9569 & 99 & & & \\
NGC~4278 & 4741, 7077 & 35, 111 & & 0205010101 & 29, 30, 24 \\
& 7078, 7079 & 52, 106 & & & \\
& 7080, 7081 & 56, 112 & & & \\
& 11269, 12124 & 78, 25 & & & \\
NGC~4365 & 2015, 5921 & 39, 38 & & 0205090101 & 25, 26, 24 \\
& 5922, 5923 & 39, 35 & & & \\
& 5924, 7224 & 25, 41 & & & \\
NGC~4374 & 803, 5908 & 27, 45 & & & \\
& 6131 & 32 & & & \\
NGC~4382 & 2016 & 40 & & 0201670101 & 19, 19, 17 \\
NGC~4406 & 318 & 32 & & 0108260201 & 73, 75, 50 \\
NGC~4459 & 2927 & 10& & 0550540101 & 74, 74, 71 \\
& 11784 & 30 & & 0550540201 & 19, 19, 17 \\
NGC~4472 & 321, 11274 & 31, 39 & & 0200130101 & 83, 83, 80 \\
NGC~4477 & 9527 & 35 & & 0112552101 & 13, 13, 9 \\
NGC~4526 & 3925 & 40 & & 0205010201 & 22, 22, 21 \\
NGC~4552 & 2072 & 51 & & 0141570101 & 24, 27, 19 \\
NGC~4636 & 323 & 43 & & 0111190701 & 59, 59, 57 \\
NGC~4649 & 785 & 16 & & 0021540201 & 49, 46, 46 \\
& 8182, 8507 & 37, 17 & & 0502160101 & 74, 74, 70 \\
NGC~4697 & 784, 4727 & 38, 40 & & 0153450101 & 47, 50, 45 \\
& 4728, 4729 & 34, 22 & & & \\
& 4730 & 35 & & & \\
NGC~5846 & 788, 7923 & 21& 87 & 0021540501 & 15, 15, 11 \\
& 7062, 8448 & 23, 8 & & 0602010101 & 85, 85, 80\\
& 8449 & 19 & & & \\
NGC~5982 & & & &0693300301 & 23, 24, 21\\
NGC~7619 & 3955 & 29 & & 0149240101 & 40, 40, 37 \\
\enddata
\end{deluxetable}

\begin{deluxetable}{lccc} 
\tablewidth{0pc}
\tabletypesize{\scriptsize}
\tablecaption{$L_{X}$, $L_{K}$, and hot gas mass}
\tablehead{
\colhead{Name} &\colhead {$L_{X}$} & \colhead{$L_{K}$}&\colhead{$M_{X,\rm gas}$} \\
\colhead{} &\colhead{ ($10^{40}$ ergs s$^{-1}$)} & \colhead{($10^{10}L_{\odot}$)}&\colhead{($10^8M_{\odot}$)}}
 \startdata
 IC~1459	         &2.1 &26.7&2.2\\
 NGC~507        & 168.7 & 56.1 &169.0\\
 NGC~720        & 4.4 & 17.0 &4.2\\
 NGC~1291     &0.3 & 7.8 &0.5\\
 NGC~1316	 & 3.7 & 43.1 &5.7\\
 NGC~1332	 & 1.0 & 14.5 &1.7\\
 NGC~1380	& 1.0 & 9.1 &3.4\\
 NGC~1395	&3.0 & 20.1 &4.1\\
 NGC~1399	&9.1 & 18.8 & 8.3\\
 NGC~1404	& 10.0 &13.4 & 4.3\\
NGC~1407       & 8.6 &  38.2& 6.1\\
NGC~3607	& 1.3 & 12.9 &2.5\\
 NGC~3608      &1.3&2.9 &3.2\\
NGC~3923	 &4.0 &	24.2 &6.7\\
 NGC~4125	 & 3.7 & 21.8 &7.5\\
NGC~4261	& 1.8 & 8.8 &7.5\\
  NGC~4278	 & 0.6 &	6.3 &1.0\\
 NGC~4365	 & 0.5 & 16.0 &1.9\\
 NGC~4374	 & 5.4 & 22.4  &4.9\\
  NGC~4382	 & 1.4 & 20.2&2.0\\
  NGC~4406	 &10.4 & 26.2 &8.7\\
  NGC~4459	 & 0.07 & 1.5 &0.5\\
  NGC~4472	& 13.7 & 35.3&10.5\\
  NGC~4477	 &1.1 & 5.7 &3.1\\
  NGC~4526	 & 0.4 & 13.0 &0.7\\
  NGC~4552	 & 2.0 & 7.7 &0.7\\
 NGC~4636	&20.6 & 10.8 &6.8\\
  NGC~4649	 & 10.6 & 29.4 &5.6\\
  NGC~4697	 & 0.5 & 8.2 &2.2\\
  NGC~5846	 & 30.5 & 23.8 &17.8\\
  NGC~5982     & 8.7 & 21.8 & 22.2 \\
 NGC~7619	 & 18.7 & 32.4 &19.3\\
 \enddata 
 \tablecomments{All quantities were measured within two effective radii given in Table~1.}
  \end{deluxetable}

 \begin{deluxetable}{lllllll}
\tablewidth{0pc}
\tabletypesize{\scriptsize}
\tablecaption{Spectral analysis results}
\tablehead{
\colhead{Name}&\colhead{Fe}&\colhead{$T_1$}&\colhead{$L_{X1}$}&\colhead{$T_2$}&\colhead{$L_{X2}$}&\colhead{$\chi^{2}/d.o.f.$}\\
\colhead{}&\colhead{}&\colhead{(keV)}&\colhead{($10^{40}$ ergs s$^{-1}$)}&\colhead{(keV)}&\colhead{($10^{40}$ ergs s$^{-1}$)}&\colhead{}\
}
\startdata
IC~1459 & $0.64^{+0.32}_{-0.27}$ &$0.31^{+0.05}_{-0.08}$ & $0.56$ & $0.99^{+0.03}_{-0.09}$&$1.57$ & 422.03/374\\
NGC~507 & $0.79^{+0.12}_{-0.11}$ & $1.00^{+0.03}_{-0.04}$&$65.4$&$1.49^{+0.09}_{-0.09}$&$105$&958.90/910\\
NGC~720 & $0.91^{+0.22}_{-0.16}$ & $0.26^{+0.08}_{-0.04}$&$0.65$&$0.60^{+0.02}_{-0.02}$&$3.76$&1354.57/1177\\
NGC~1291 & $0.28^{+0.39}_{-0.15}$ & $0.20^{+0.03}_{-0.04}$&$0.130$&$0.91^{+0.11}_{-0.15}$&$0.12$&789.78/788\\
NGC~1316 & $0.51^{+0.04}_{-0.04}$ & $0.52^{+0.10}_{-0.03}$&$1.77$&$0.95^{+0.03}_{-0.03}$&$1.93$&2114.24/1780\\
NGC~1332 & $0.29^{+0.21}_{-0.08}$ & $0.39^{+0.10}_{-0.03}$&$1.03$&$1.73^{+0.78}_{-0.40}$&$1.03$&143.91/139\\
NGC~1380 & $0.50^{+0.65}_{-0.23}$ &$0.21^{+0.03}_{-0.03}$&$0.73$&$0.54^{+0.06}_{-0.09}$&$0.66$&231.54/221\\
NGC~1395 & $0.76^{+0.19}_{-0.13}$ &$0.68^{+0.06}_{-0.06}$&$0.69$&$0.82^{+0.84}_{-0.06}$&$2.32$&569.80/533\\
NGC~1399 & $0.78^{+0.04}_{-0.04}$ &$0.88^{+0.09}_{-0.02}$&$\sim 0$&$1.04^{+0.01}_{-0.01}$&$9.10$&2123.75/1351\\
NGC~1404 & $0.70^{+0.08}_{-0.05}$ &$0.34^{+0.03}_{-0.01}$&$3.83$&$0.69^{+0.03}_{-0.02}$&$6.19$&619.60/467\\
NGC~1407 & $1.98^{+1.02}_{-0.53}$ & $0.83^{+0.02}_{-0.02}$&$7.35$&$1.69^{+0.41}_{-0.14}$&$1.21$&791.31/719\\
NGC~3607 & $0.29^{+0.13}_{-0.07}$ & $0.43^{+0.09}_{-0.06}$&$1.11$&$0.63^{+0.20}_{-0.04}$&$0.06$&238.33/200\\
NGC~3608 & $1.02^{+1.10}_{-0.43}$ & $0.19^{+0.03}_{-0.03}$&$0.09$&$0.54^{+0.04}_{-0.09}$&$0.38$&319.85/294\\
NGC~3923 & $0.82^{+0.16}_{-0.12}$ & $0.30^{+0.03}_{-0.05}$&$1.51$&$0.63^{+0.03}_{-0.02}$&$2.46$&1470.81/1330\\
NGC~4125 & $0.52^{+0.46}_{-0.18}$ & $0.25^{+0.07}_{-0.04}$&$1.81$&$0.54^{+0.10}_{-0.07}$&$1.87$&166.83/176\\
NGC~4261 & $0.58^{+0.48}_{-0.21}$ & $0.72^{+0.07}_{-0.07}$&$1.39$&$3.02^{+1.55}_{-1.35}$&$0.47$&200.54/198\\
NGC~4278 & $0.39^{+0.68}_{-0.23}$ & $0.08^{+0.06}_{-0.08}$&$0.41$&$0.74^{+0.10}_{-0.07}$&$0.21$&452.50/444\\
NGC~4365 & $0.56^{+1.27}_{-0.23}$ & $0.25^{+0.07}_{-0.04}$&$\sim 0$&$0.58^{+0.04}_{-0.06}$&$0.45$&806.15/783\\
NGC~4374 & $0.60^{+0.26}_{-0.14}$ & $0.27^{+0.15}_{-0.27}$&$5.4$&$0.69^{+0.04}_{-0.06}$&$\sim 0$&499.16/496\\
NGC~4382 & $0.62^{+0.26}_{-0.18}$ & $0.29^{+0.04}_{-0.04}$&$1.16$&$0.75^{+0.10}_{-0.33}$&$0.24$&305.38/298\\
NGC~4406 & $0.63^{+0.03}_{-0.03}$ & $0.77^{+0.01}_{-0.05}$&$10.4$&$0.94^{+0.01}_{-0.02}$&$0.1$&1564.47/1169\\
NGC~4459 & $0.22^{+0.09}_{-0.06}$ & $0.52^{+0.07}_{-0.06}$&$0.07$&$2.94^{+1.38}_{-0.95}$&$\sim 0$&433.43/429\\
NGC~4472 & $0.79^{+0.03}_{-0.04}$ & $0.95^{+0.01}_{-0.01}$&$11.9$&$1.67^{+0.06}_{-0.10}$&$1.80$&1528.00/1189\\
NGC~4477 & $0.64^{+0.39}_{-0.21}$ & $0.24^{+0.02}_{-0.03}$&$0.75$&$0.67^{+0.03}_{-0.10}$&$0.38$&228.77/194\\
NGC~4526 & $0.60^{+1.09}_{-0.32}$ & $0.23^{+0.02}_{-0.02}$&$0.36$&$0.54^{+0.68}_{-0.32}$&$0.06$&205.81/212\\
NGC~4552 & $0.54^{+0.15}_{-0.10}$ & $0.39^{+0.06}_{-0.06}$&$1.05$&$0.70^{+0.05}_{-0.05}$&$1.00$&523.57/468\\
NGC~4636 & $1.00^{+0.05}_{-0.04}$ & $0.31^{+0.01}_{-0.01}$&$1.55$&$0.75^{+0.01}_{-0.01}$&$19.1$&1926.00/1117\\
NGC~4649 & $1.06^{+0.01}_{-0.01}$ & $0.86^{+0.01}_{-0.01}$&$10.6$&$1.03^{+0.02}_{-0.02}$&$\sim 0$&3700.31/2301\\
NGC~4697 & $0.42^{+0.22}_{-0.14}$ & $0.21^{+0.02}_{-0.08}$&$0.30$&$0.40^{+0.11}_{-0.08}$&$0.16$&1026.25/1017\\
NGC~5846 & $0.79^{+0.07}_{-0.07}$ & $0.44^{+0.08}_{-0.05}$&$5.08$&$0.76^{+0.03}_{-0.02}$&$25.4$&1015.02/817\\
NGC~7619 & $1.06^{+0.21}_{-0.18}$ & $0.85^{+0.02}_{-0.02}$&$18.3$&$0.88^{+0.29}_{-0.03}$&$0.535$&448.73/420\\
\enddata   
\tablecomments{T$_1$ (L$_1$) and T$_2$ (L$_2$) represent the best fit temperatures (luminosities) of each {\tt vapec} component in the two temperature model. All quantities were measured within two effective radii given in Table~1.}
 \end{deluxetable}

\begin{deluxetable}{lccc}
\tabletypesize{\scriptsize}
\tablewidth{0pc}
\tablecaption{Summary of correlations for each relation}
\tablehead{
\colhead{Relations}&\colhead{Correlation coefficient $\rho$}&\colhead{Null hypothesis probability}&\colhead{Number of galaxies}}
\startdata
Fe--$L_X/L_K$&0.449&0.0124&32\\
Fe--$L_X/L_K$&0.673&0.0012&24$^{\ast}$\\
Fe--Age&-0.179&0.3193&32\\
Fe--$Z_{\rm star}$&0.146&0.5032&22\\
Fe--M(H$_2$)&-0.409&0.0498&24$^{\ast\ast}$\\
Fe--M(H~{\sc i}) &-0.089&0.6709&24$^{\ast\ast}$\\
Fe--M(H$_2$)/M(X)&-0.459&0.0276&24$^{\ast\ast}$\\
Fe--M(H~{\sc i})/M(X)&-0.059&0.7769&24$^{\ast\ast}$\\
\hline
M(H$_2$)--M(X)&-0.323&0.1211&24$^{\ast\ast}$\\
M(H~{\sc i})--M(X)&-0.335&0.1085&24$^{\ast\ast}$\\
\enddata
\tablecomments{$^{\ast}$ group center galaxies excluded.
$^{\ast\ast}$ galaxies with cold gas data.}
 \end{deluxetable}

 \begin{figure}
 \epsscale{0.7}

 \plotone{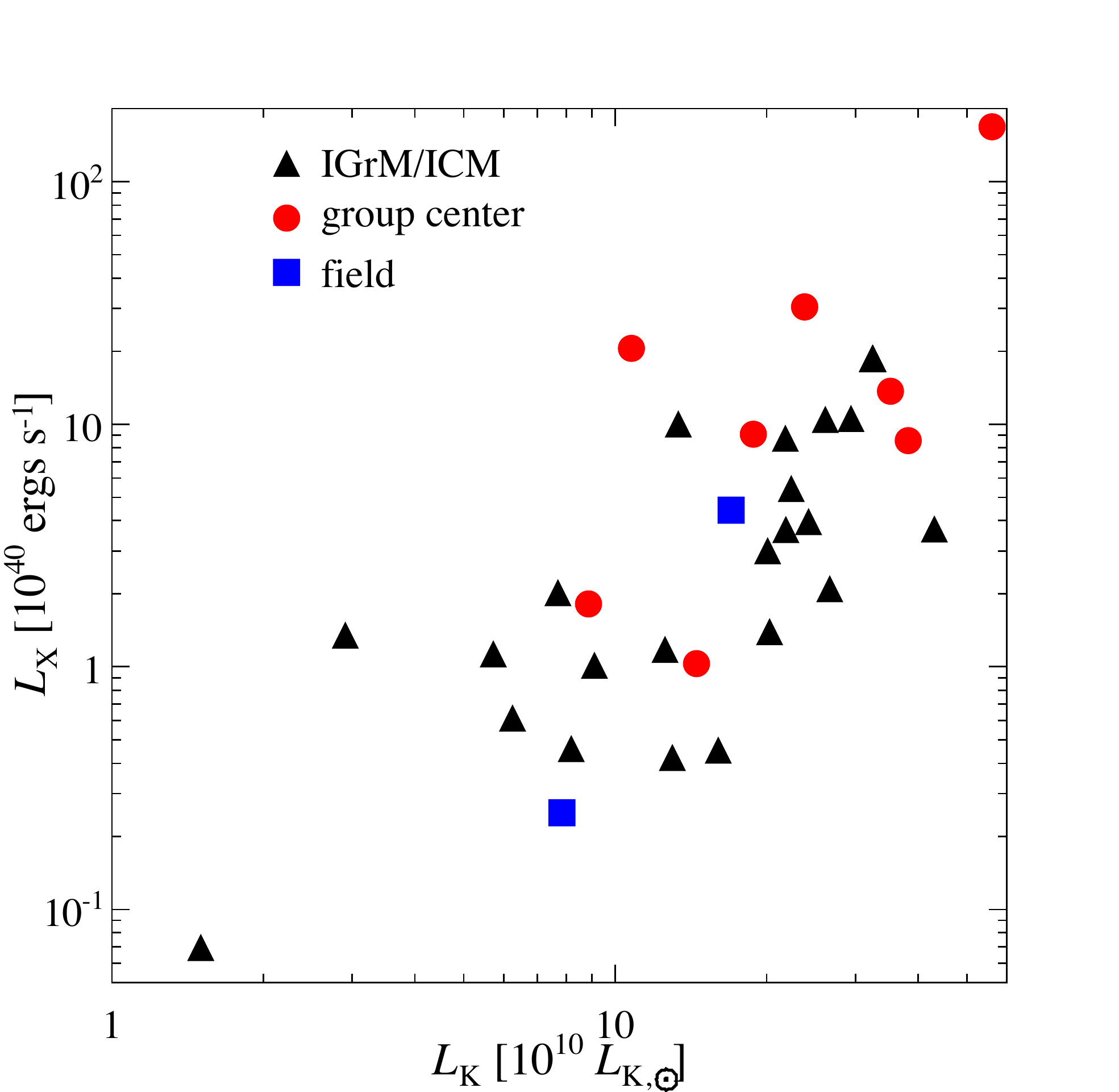}
  \caption{$L_X$ as a function of $L_{K}$. Circles: galaxies at group centers. Triangles: galaxies in groups/clusters but not at centers. Boxes: galaxies in the field. [{\sl see the electronic edition of the journal for a color version of this figure.}]}
\end{figure}
 
 \begin{figure}

  \plotone{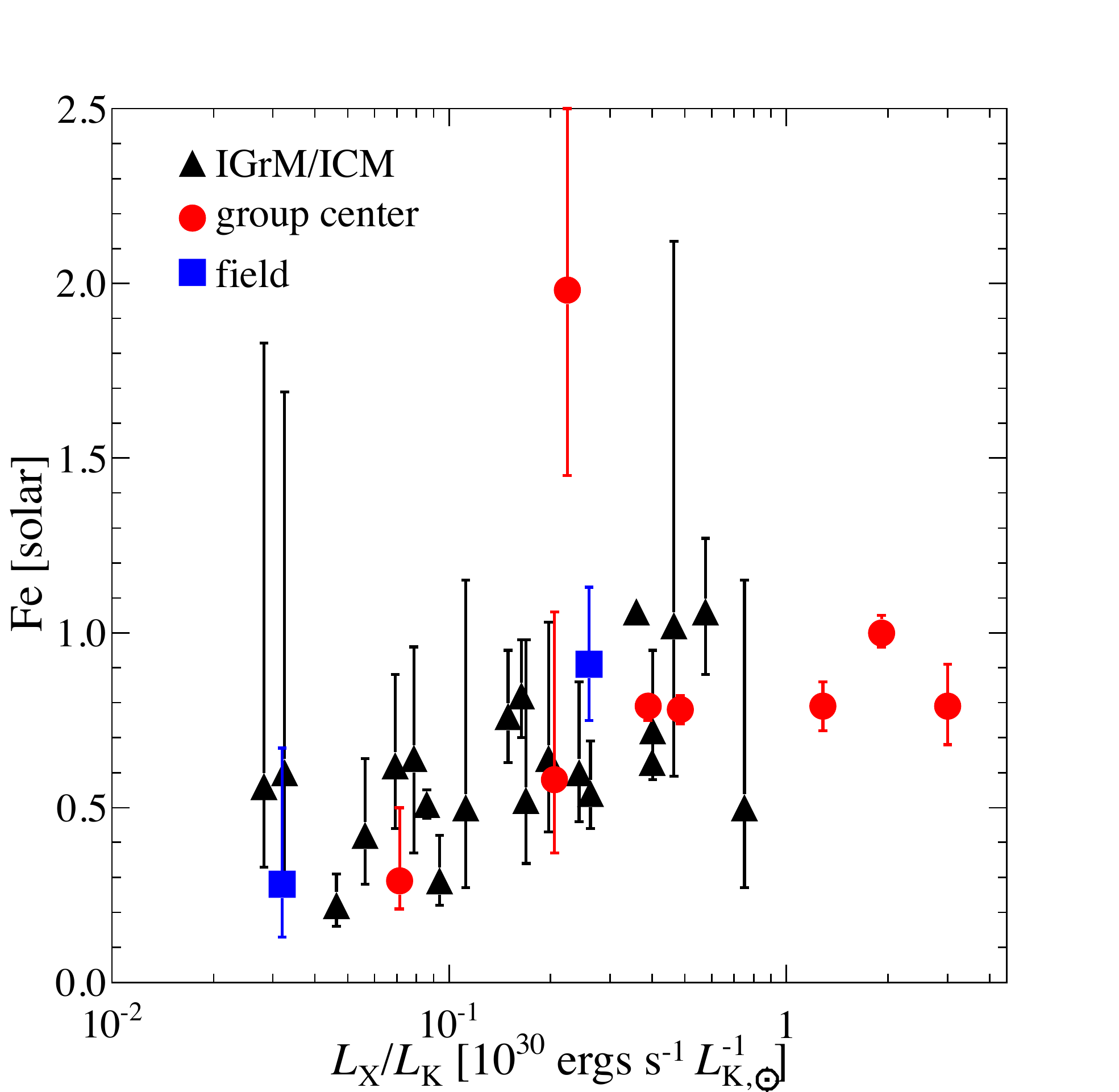}
  \caption{Best fit hot gas Fe abundance as a function of the ratio of $L_X$ to $L_K$. Circles: galaxies at group centers. Triangles: galaxies in groups/clusters but not at centers. Boxes: galaxies in the field. [{\sl see the electronic edition of the journal for a color version of this figure.}]}
\end{figure}

 \begin{figure}
  \epsscale{1.2}
 \plottwo{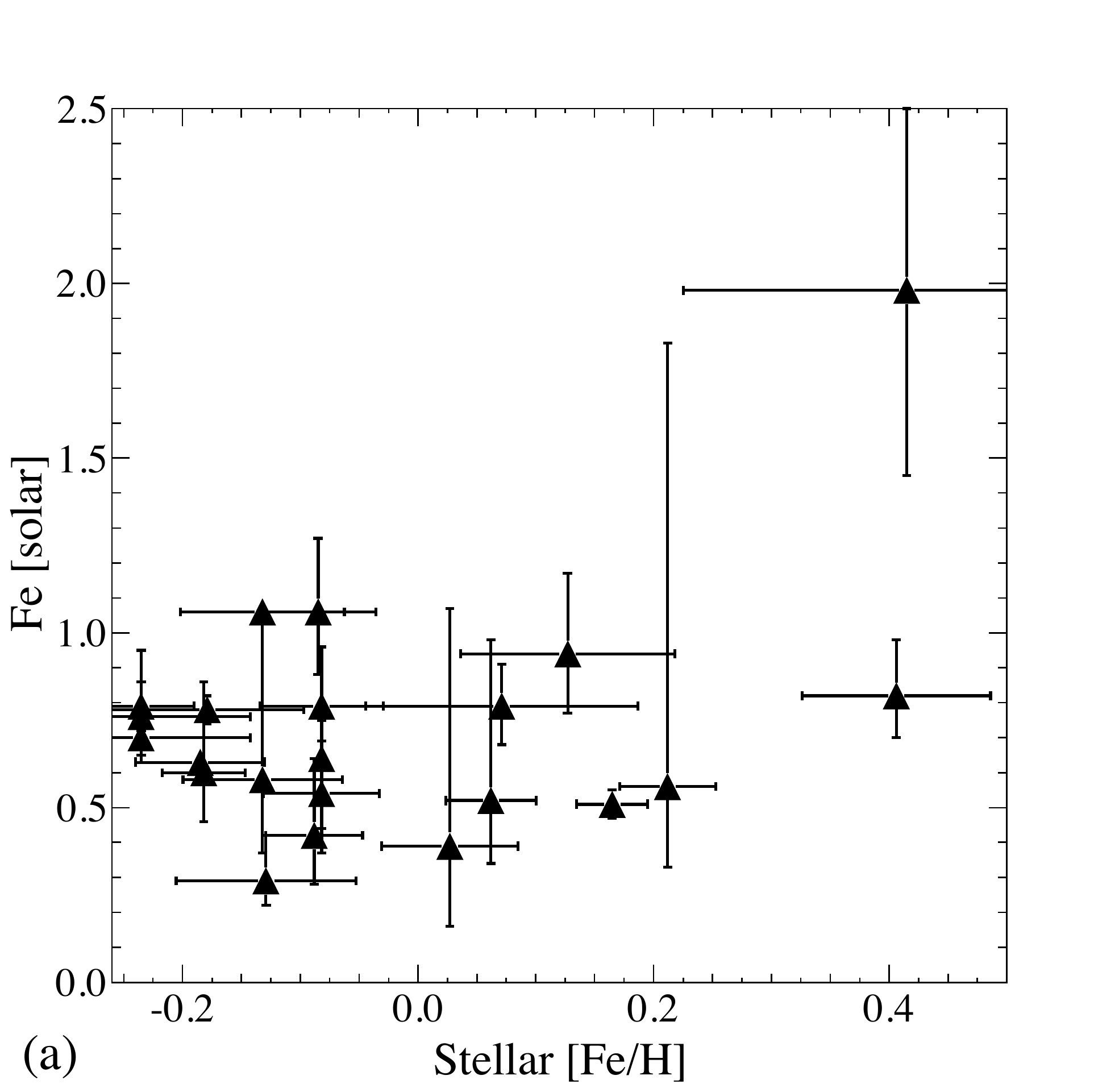}{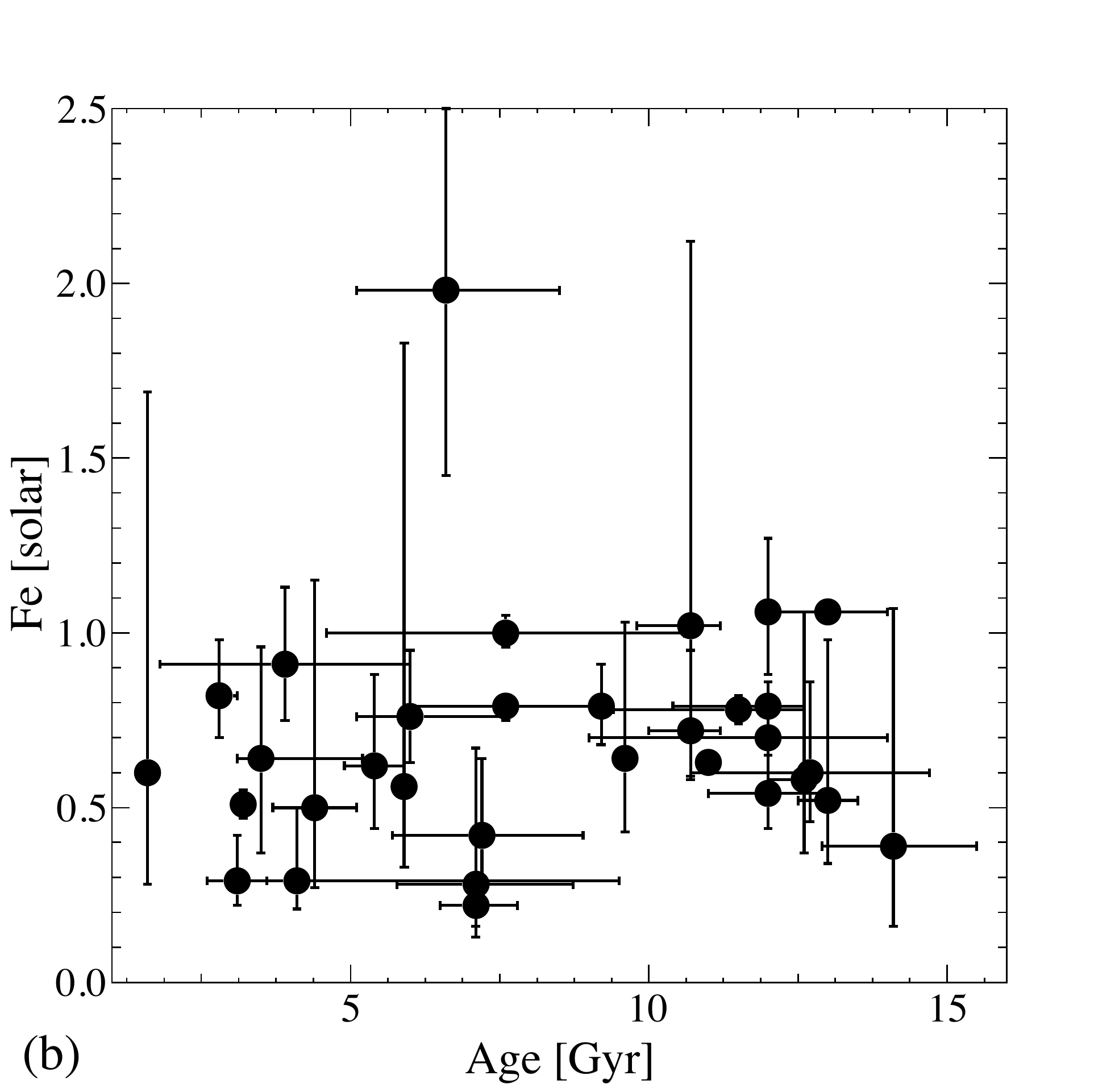} 
  \caption{(a): best fit hot gas Fe abundance as a function of stellar Fe abundance. (b): best fit hot gas Fe abundance as a function of stellar age. Hot gas Fe abundance is not correlated with stellar Fe abundance or stellar age. }

\end{figure}

 \begin{figure}
  \epsscale{1.2}

 \plottwo{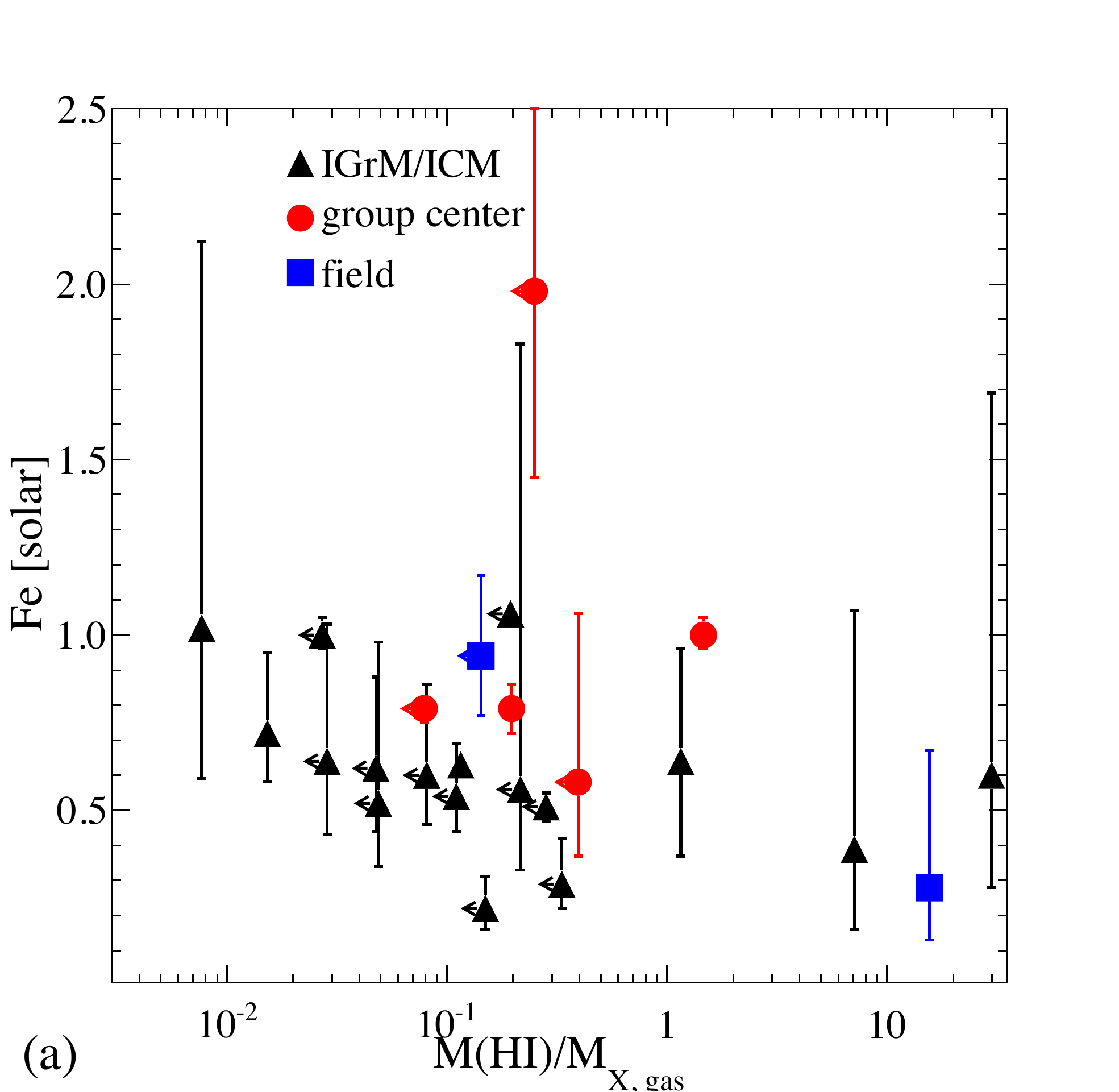}{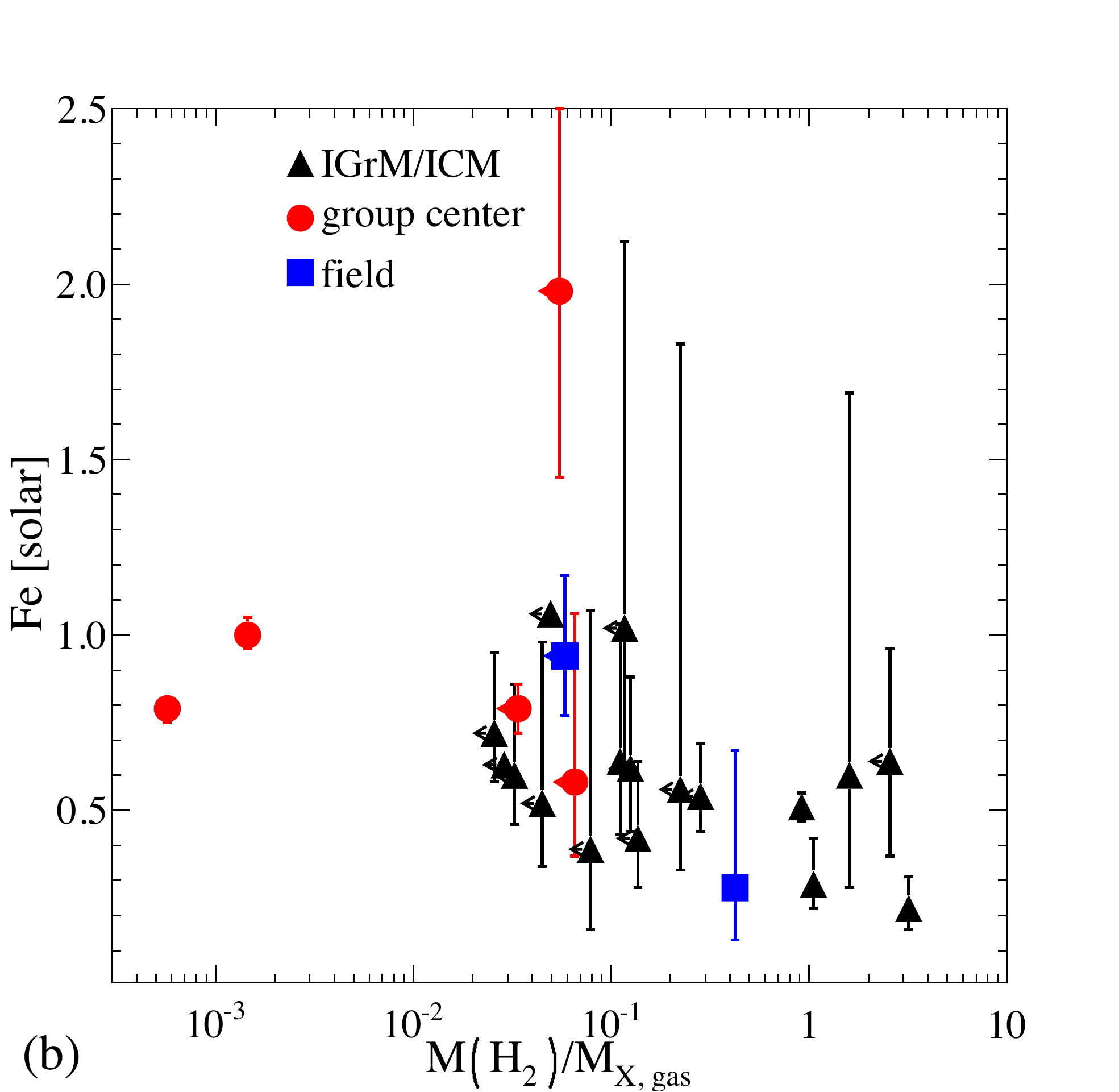}
  \caption{(a): The best fit hot gas Fe abundance as a function of the ratio of atomic gas mass $M_{HI}$ to hot gas mass $M_{X, \rm gas}$. (b): The best fit hot gas Fe abundance as a function of the ratio of molecular gas mass $M_{H_2}$ to hot gas mass $M_{X, \rm gas}$. 
Circles: galaxies at group centers. Triangles: galaxies in groups or clusters but not at group centers. Boxes: galaxies in the field. One sided error bars represent upper limit on cold gas mass.
 [{\sl see the electronic edition of the journal for a color version of this figure.}]}
\end{figure}

 \begin{figure}
 \epsscale{0.7}
  \plotone{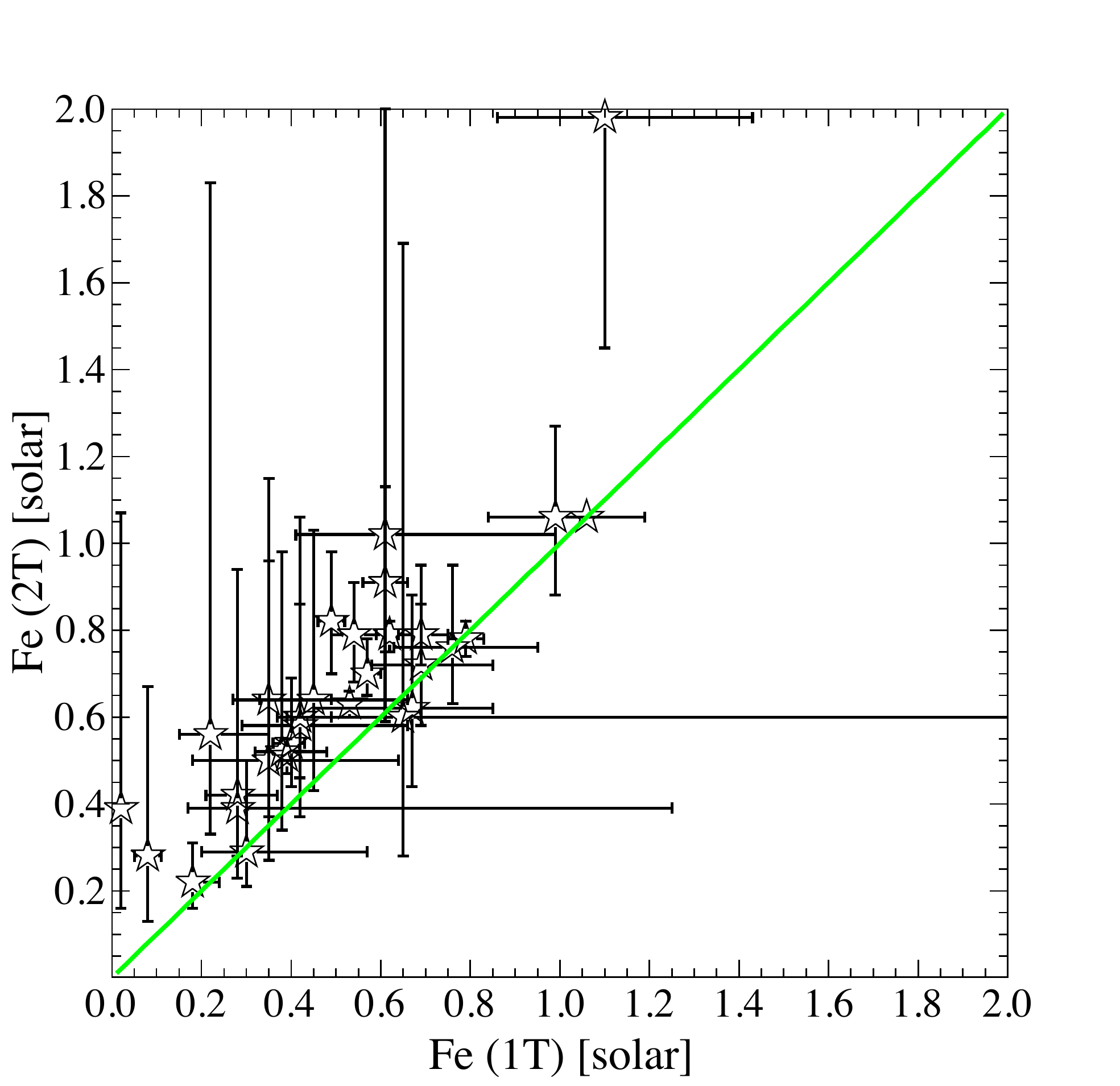}
  \caption{ comparison of best fit Fe abundance obtained with one thermal temperature model versus two thermal temperature model. Solid line: Fe (1T) = Fe (2T). [{\sl see the electronic edition of the journal for a color version of this figure.}]}
\end{figure}

\begin{figure}
 \epsscale{1.2}

 \plottwo{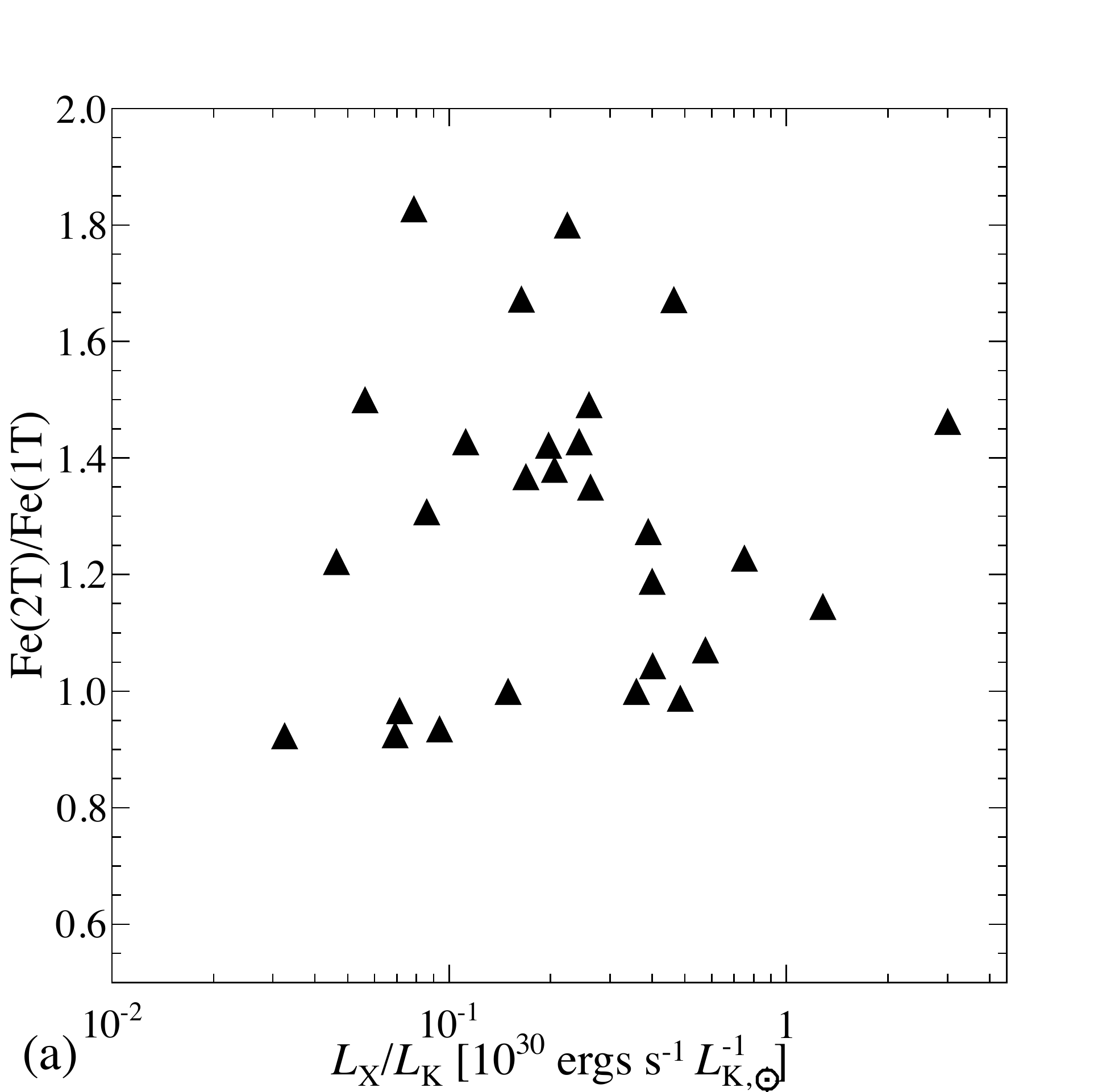}{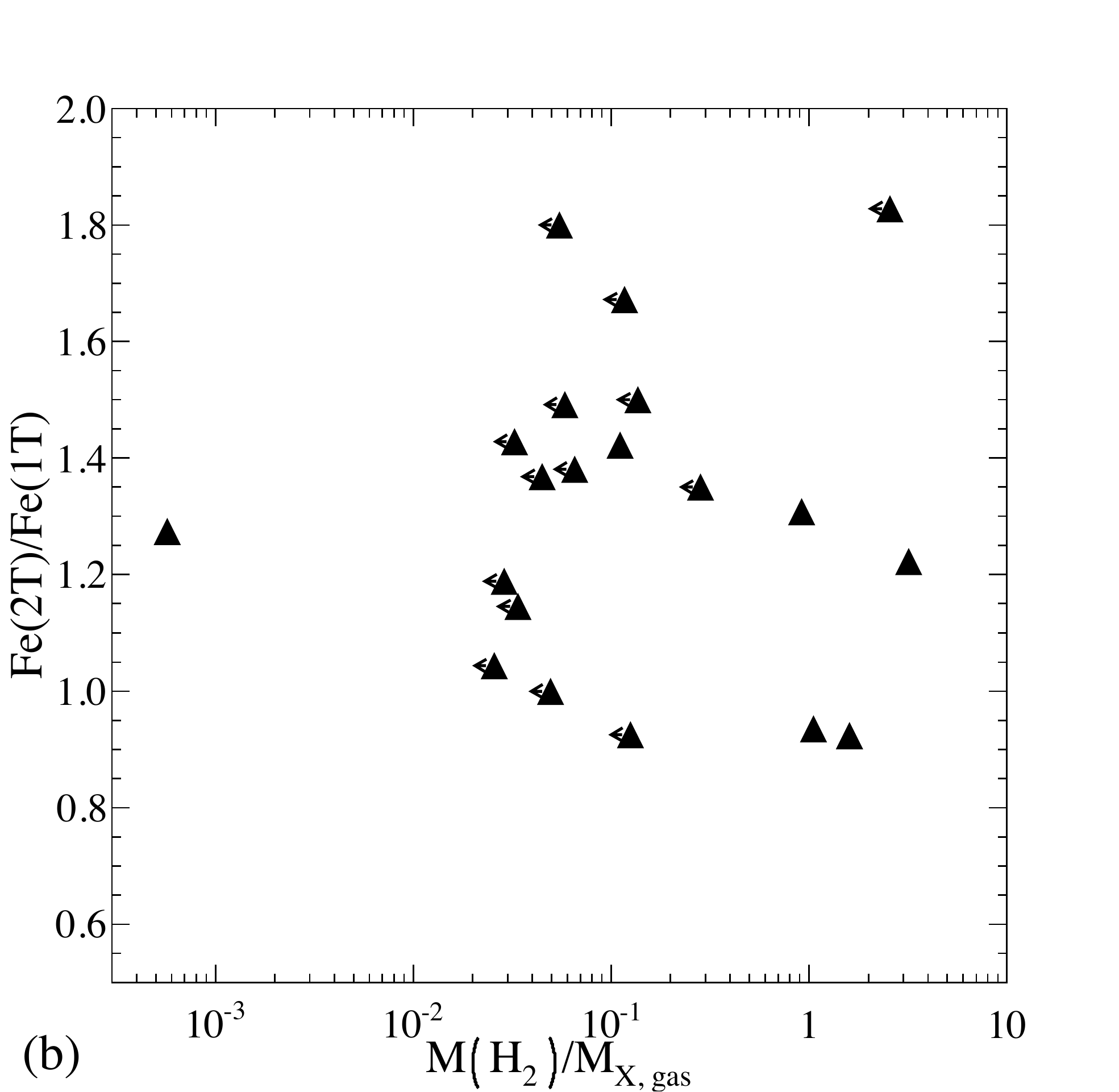}
  \caption{(a): the ratio of Fe (2T) of Fe (1T) as a function of the ratio of $L_X$ to $L_K$. (b): the ratio of Fe (2T) of Fe (1T) as a function of the ratio of M($H_2$) to M$_{X,\rm gas}$.}

\end{figure}

\end{document}